\documentclass{article}

\usepackage[preprint]{neurips_2025}

\usepackage[utf8]{inputenc} 
\usepackage[T1]{fontenc}    
\usepackage{hyperref}       
\usepackage{url}            
\usepackage{booktabs}       
\usepackage{amsfonts}       
\usepackage{nicefrac}       
\usepackage{microtype}
\usepackage{soul}
\usepackage[table]{xcolor}
\usepackage{enumitem}
\usepackage{lipsum}       
\usepackage{changepage}   
\usepackage[symbol]{footmisc}

\usepackage{graphicx}  
\usepackage{float}     
\usepackage{booktabs}  
\usepackage{caption}   
\usepackage{subcaption} 
\usepackage{float}
\usepackage{hyperref}  
\hypersetup{
    colorlinks=true,
    linkcolor=blue,
   citecolor=blue,
    urlcolor=blue
}
\usepackage{amsmath}
\usepackage{algorithm}
\usepackage{algpseudocode}
\usepackage{textcomp}
\usepackage{array}

\title{Evaluating the Energy-Efficiency of the Code Generated by LLMs}

\author{
Md Arman Islam\footnote[2]{} \footnote[1]{}\thanks{Equal contribution}, Devi Varaprasad Jonnala\footnote[2]{} \footnote[1]{}, Ritika Rekhi\footnote[2]{}\\
{\bf Pratik Pokharel\footnote[2]{}, Siddharth Cilamkoti\footnote[2]{}, Asif Imran\footnote[2]{}, Tevfik Kosar\footnote[2]{}, and Bekir Turkkan\footnote[3]{}} \\
 \footnote[2]{} University at Buffalo, Buffalo, NY, USA \\
 \footnote[3]{} IBM Research, Yorktown Heights, NY, USA
}

\begin{document}

\maketitle

\vspace{-2mm}
\begin{abstract}

As the quality of code generated by Large Language Models (LLMs) improves, their adoption in the software industry for automated code generation continues to grow. Researchers primarily focus on enhancing the functional correctness of the generated code while commonly overlooking its energy efficiency and environmental impact. This paper investigates the energy efficiency of the code generated by 20 popular LLMs for 878 programming problems of varying difficulty levels and diverse algorithmic categories selected from the LeetCode platform by comparing them against canonical human-written solutions. Although LLMs can produce functionally correct results in most cases, our findings show that the performance and energy efficiency of LLM-produced solutions are often far below those of human-written solutions. Among the studied LLMs, DeepSeek-v3 and GPT-4o generate the most energy-efficient code, whereas Grok-2 and Gemini-1.5-Pro are among the least energy-efficient models. On average, human-generated canonical solutions are approximately {\bf 1.17} times more energy efficient than DeepSeek-v3, {\bf 1.21} times more energy efficient than GPT-4o, and over {\bf 2} times more energy efficient than Grok-2 and Gemini-1.5-Pro. For specific algorithmic groups such as dynamic programming, backtracking, and bit manipulation, LLM-generated code can consume up to {\bf 450} times more energy than human-generated canonical solutions.
%
%

\end{abstract}

\vspace{-3mm}
\section{Introduction}
\vspace{-1mm}

In software development, AI-assisted code generators have become vital to increase productivity, maintain consistency, enforce standards, and refine existing codebases. Industry leaders are increasingly adopting LLMs to automate the process of code generation, testing, and project document writing. 
Similarly, LLMs help developers create code structures based on specifications, enhance development efficiency, and decrease the likelihood of errors.

Traditionally, evaluations of code generators have centered around runtime efficiency and code quality, while they commonly ignore the energy consumption and environmental impact of generated code. 
However, evolutions in Generative AI and the increasing demand for Information Technology (IT) software systems have caused emerging sustainability issues, particularly due to generative models \cite{vartziotis2024learn}.
The IT industry contributes about 10\% of global energy use today \cite{ict_trends_erol_2023} and is responsible for approximately 3\% of global carbon emissions\cite{columbia2023ai}, exceeding the emissions of the aviation industry \cite{atag2023facts}. Training a large neural network can cause over 626,000 pounds of CO\textsubscript{2} emissions, nearly five times the lifetime emissions of an average car~\cite{patterson2021carbon}.
The demand for computing power, mainly due to technologies like Generative AI, is expected to cause data centers to consume 20\% of global electricity by 2030 \cite{belkhir2018assessing}. 
On the other hand, improving these power-hungry models regarding energy efficiency can reduce the high power demand. 
Li et al. \cite{li2023green} show that transformations in code execution can enable savings in cloud computing costs of around 42 percent without losing any functionality. 


This paper fills an important gap in this area by investigating the energy efficiency of code generated by LLMs and comparing it to canonical human-written solutions.
We systematically evaluate the energy efficiency of code snippets generated by 20 popular LLMs for 878 programming problems with different difficulty levels selected from EffiBench~\cite{huang2024effibench}. We compare generated code with human-generated canonical solutions to find patterns by measuring and evaluating the energy consumption. This study allows us to gain insights into the environmental impact and economic cost of using LLM-generated code. 
%
We highlight the importance of sustainable models in code generation. By raising awareness of the energy costs associated with AI-assisted code generation, we aim to encourage the development of AI tools that produce environmentally sustainable code.

Among the studied LLMs, DeepSeek-v3, GPT-4o, and Claude-3.5-Sonnet generate the most energy-efficient code in general, whereas Llama-3.3-70B, Grok-2, and Gemini-1.5-Pro are among the least energy-efficient models.
On average, human-generated canonical solutions were approximately {\bf 1.17} times more energy efficient than DeepSeek-v3, {\bf 1.2} times more energy efficient than GPT-4o and Claude-3.5-Sonnet, {\bf 1.93} times more energy efficient than Llama-3.3-70B, and over {\bf 2} times more energy efficient than Grok-2 and Gemini-1.5-Pro. 
For specific algorithmic groups such as
dynamic programming, backtracking, and bit manipulation, the LLM-generated
code can consume significantly more energy compared to human-generated canonical solutions.
For these problem categories, GPT-4o generates solutions consuming up to {\bf 46} times more energy than the canonical solution. Similarly, LLaMA-3.3-70B generates solutions with energy consumptions up to {\bf 149} times that of the canonical solution, and Gemini-1.5-Pro generates solutions that consume energy up to {\bf 449} times that of the canonical solution.
%
%
%
These results suggest that energy efficiency optimization should become an important consideration in the development of next-generation AI-assisted code generation systems. 

In summary, the paper makes the following contributions:
\begin{adjustwidth}{0.6cm}{}

\vspace{-1mm}
\textbf{(1)} We conduct an extensive evaluation of 20 LLMs, and compare the energy efficiency of the code generated by them.

\vspace{-1mm}
\textbf{(2)} We ensure fair prompt inputs during code generation and fair comparison of LLMs against each other and canonical solutions by considering problems correctly generated by all models.

\vspace{-1mm}
\textbf{(3)} We propose a comprehensive evaluation framework that jointly measures energy consumption, runtime performance, memory usage, the number of input and output tokens used for code generation, and the monetary cost of generating the code, providing a holistic understanding of the environmental and economic costs associated with LLM-generated code.


\vspace{-1mm}
\textbf{(4)} Our findings reveal that while advanced LLMs like DeepSeek-v3, GPT-4o, and Claude-3.5-Sonnet can produce more efficient code than other models, they still are considerably less efficient than human-written solutions in terms of energy consumption.

\vspace{-1mm}
\textbf{(5)} Our analysis identifies clear patterns in energy inefficiency, showing that LLMs particularly struggle with Dynamic Programming, Backtracking, Bit Manipulation, and Greedy algorithms; while performing relatively better on problems involving Binary Search and Divide and Conquer. 

\end{adjustwidth}

\section{Related Works}

{\bf Green Software Practices:} The idea of green software practices relates to energy consumption and sustainability in the end-to-end software development lifecycle \cite{penzenstadler2014green,verdecchia2022green}. Green software examples include energy-aware coding and sustainable development methodologies \cite{ahmed2017green, kaestner2013energy}. 
Sustainability requires consideration from the start of all aspects of development, rather than an after-thought \cite{verdecchia2022green}. Clean code methodology can help to execute with better efficiencies \cite{ikegwu2024energy}, and it can focus on reducing instructions, removing duplicate code, and optimizing an algorithm \cite{rashid2021advancing}. 
Additional suggestions to reduce energy use include parallelization, caching, and compression \cite{devsu2024sustainable}. Henderson et al. \cite{henderson2020towards} offer standardized energy reporting to allow others' projects to replicate it. The ESC framework provides a consistent sustainability computing paradigm to take a holistic approach \cite{patil2023holistic}. Addressing the carbon footprint of computation, the study on "Green Algorithms" \citep{lannelongue2020green} introduces a quantitative model for assessing the environmental impact of computational processes.


{\bf Efficiency of LLM-generated Code:} When examining efficiency for LLM-generated code, models should consider correctness, memory, runtime, and energy \cite{shi2024enamel}. Chen et al. \cite{chen2023neural} propose developing multi-faceted assessments. EffiBench \cite{huang2024effibench}
constructs an LLM code efficiency benchmark with 1000 coding problems representing different algorithmic complexities. Cruz et al. \cite{cruz2023functional} state that efficiency can only be evaluated after correctness is established, whereas Niu et al. \cite{niu2024evaluating} argue that efficiency depends neither on correctness nor model size, but efficiency does scale up through prompting piecewise. 
%
LLMs have a well-documented issue with selecting unoptimized algorithms and iterations and data structures \cite{koutsouris2024unveiling}; perhaps the most obvious comparison is QuickSort vs InsertionSort regarding time complexity \cite{koutsouris2024unveiling}. Energy-aware prompting could make a 30 percent reduction \cite{lu2023energy}, and subsequent feedback loops using evaluator LLMs could identify further means for optimizing the code generated \cite{bilal2024towards}. In green code, the system produces emission reductions of 23 to 50 percent for generation tasks, through reinforcement learning, rewarding the reduction of emissions \cite{ilager2025green}. Wang et al. \cite{wang2024reinforcement} study the effect of providing energy rewards. There is also emerging research interest in the utility of prompt engineering for reduced energy use, though with less clearly predictable results so far \cite{cappendijk2024generating}. Vartziotis et al. \cite{vartziotis2024learn} define "green capacity" to capture sustainability in AI-generated code.

\section{Benchmarking Approach and Experiments}

To construct our benchmark for comparing different LLMs in terms of their energy efficiency, we use an approach similar to EffiBench~\cite{huang2024effibench}, which is inspired by the common practice of evaluating developers’ coding ability using problems from the competitive coding platform – LeetCode~\cite{leetcode}. EffiBench includes 1000 Leetcode problems that are asked in interviews frequently (>40\%). These problems are paired with the most efficient solutions from the LeetCode discussion forum, labeled as the canonical human-written solutions. 100 test cases for each problem are included in EffiBench which are generated using a test case generator based on GPT-3.5-turbo. 

\subsection{Problem Dataset Selection}

Before using the EffiBench dataset for our study, we thoroughly analyzed and tested the dataset. The following are the findings from our analysis: (i) 12 problems in the dataset do not have comprehensive test cases; (ii) 110 problems throw errors when we run canonical solutions against comprehensive test cases. This is due to one or all of the following reasons: syntax errors in canonical solutions, syntax errors in comprehensive test cases, and improper definition of test cases for TreeNode, GraphNode, and LinkedList problems.
We excluded these 122 problems from our dataset and considered 878 problems for
our comprehensive study. 

\begin{table}[H]
\vspace{-3mm}
    \caption{Algorithm categories and difficulty-wise problem distribution in the selected dataset.}
    \label{table:benchmark_statistics}
    \centering
    \scriptsize 
    \resizebox{\textwidth}{!}{%
    \begin{tabular}{l c c c >{\centering\arraybackslash}p{1.2cm} c c >{\centering\arraybackslash}p{1.2cm} >{\centering\arraybackslash}p{1.2cm} >{\centering\arraybackslash}p{1.2cm} >{\centering\arraybackslash}p{1cm} c c}
        \toprule
        \textbf{Algorithm} & \textbf{Greedy} & \textbf{DP} & \textbf{Backtrack} & \textbf{Divide \& Conquer} & \textbf{DFS} & \textbf{BFS} & \textbf{Binary Search} & \textbf{Two Pointers} & \textbf{Sliding Window} & \textbf{Bit Manip.} & \textbf{Sorting} & \textbf{TOTAL} \\
        \midrule
        \textbf{Easy} & 32 & 7 & 1 & 4 & 3 & 0 & 23 & 31 & 8 & 25 & 62 & \cellcolor{gray!25}145 \\
        \textbf{Medium} & 161 & 140 & 31 & 6 & 37 & 36 & 68 & 52 & 40 & 54 & 128 & \cellcolor{gray!25}510 \\
        \textbf{Hard} & 38 & 111 & 8 & 9 & 13 & 22 & 48 & 5 & 14 & 18 & 40 & \cellcolor{gray!25}223 \\
        \textbf{TOTAL} & \cellcolor{gray!25}231 & \cellcolor{gray!25}258 & \cellcolor{gray!25}40 & \cellcolor{gray!25}19 & \cellcolor{gray!25}53 & \cellcolor{gray!25}58 & \cellcolor{gray!25}139 & \cellcolor{gray!25}88 & \cellcolor{gray!25}62 & \cellcolor{gray!25}97 & \cellcolor{gray!25}230 & \cellcolor{gray!25}878 \\
                \bottomrule
    \end{tabular}%
    }
\end{table}

Table~\ref{table:benchmark_statistics} shows the detailed breakdown of the 878 problems we use to compare the energy efficiency of the LLMs in our study. There are 145 easy, 510 medium, and 223 hard problems in the dataset. Leetcode defines easy, medium, and hard problems based on the complexity of the algorithms or data structures required to solve the problems. 
%
The algorithmic methods include Greedy, Dynamic Programming (DP), Backtracking, Divide and Conquer, Depth-First Search (DFS), Breadth-First Search (BFS), Binary Search, Two Pointers, Sliding Window, Bit Manipulation, and Sorting. The diverse set of algorithmic methods in the problem set provides a fair comparison of the studied LLMs across multiple problem subcategories with different computational complexities.
In the table, one problem may be tagged to more than one algorithmic category and hence the sum of the number of problems across different algorithmic categories for a given difficulty level may be greater than the reported total.

\subsection{LLMs Under Study}
We analyze 20 popular LLMs that are widely used by developers for code generation tasks. In our study, we choose 7 open-source models (from DeepSeek, Meta, and Mistral), and 13 closed-source models (from Amazon, Antropic, Google, OpenAI, and xAI). The selected models are listed in Table \ref{tab:llm-table}.
The table also shows the access type and cost of using each model -- based on the cost per 1 Million input tokens used and the cost per 1 Million output tokens generated. Some closed-source models such as GPT-4 Turbo and Claude 3.5 Sonnet incur significantly higher token processing costs, with input/output costs reaching up to \$5/\$15 per Million tokens. In contrast, open-source models such as  Llama are less expensive to access with input/output costs as low as \$0.05/\$0.08 per Million tokens. Among all LLMs studied, the least expensive one is Nova-micro, with input/output costs of \$0.02/\$0.07 per Million tokens. 
 The input/output token cost information is used in our study to compare the average cost of using each model to generate the correct code for the set of given problems. 
 The cost of open-source LLMs was determined based on their publicly available API pricing, without accounting for any additional hardware or infrastructure costs. Specifically, the models were accessed via third-party platforms such as \texttt{Fireworks.ai} and \texttt{Groq Cloud}, where the API usage charges directly reflect the input and output token processing costs. Since these experiments did not run the models on private hardware or owned cloud servers, no supplementary hardware-related expenses were included in the cost calculations.

\begin{table}[H]
    \caption{List of LLMs included in our study, their access types, and cost information (green color highlights the lowest cost, whereas red color highlights the highest cost).}
    \label{tab:llm-table}
    \centering
    \scriptsize 
    \resizebox{\textwidth}{!}{%
    \begin{tabular}{l l >{\centering\arraybackslash}p{1cm} c > {\centering\arraybackslash}p{1.8cm} >{\centering\arraybackslash}p{1.8cm}}
        \toprule
         \textbf{Vendor} & \textbf{LLM} & \textbf{Source Code} & \textbf{Access Type} & \textbf{Cost per 1M Input Tokens} & \textbf{Cost per 1M Output Tokens}\\
        \midrule
        Amazon & Nova-Lite & Closed & AWS Bedrock Batch Inference & \$0.03 & \$0.12 \\
        Amazon & Nova-Micro & Closed & AWS Bedrock Batch Inference & \cellcolor{green!40}\$0.02 & \cellcolor{green!40}\$0.07 \\
        Amazon & Nova-Pro & Closed & AWS Bedrock Batch Inference & \$0.40 & \$1.60 \\
        Anthropic & Claude 3.5 Haiku & Closed & API & \$0.80 & \$4.00 \\
        Anthropic & Claude 3.5 Sonnet & Closed & API & \$3.00 & \cellcolor{red!30}\$15.00 \\
        DeepSeek & DeepSeek v3 (37B) & Open & Fireworks.ai API & \$0.90 & \$0.90 \\
        Google & Gemini 1.5 Flash & Closed & Vertex AI Batch Prediction & \$0.04 & \$0.15 \\
        Google & Gemini 1.5 Pro & Closed & Vertex AI Batch Prediction & \$0.63 & \$2.50 \\
        Google & Gemini 2.0 Flash & Closed & Vertex AI Batch Prediction & \$0.08 & \$0.30 \\
        Google & Gemini 2.0 Flash-Lite & Closed & Vertex AI Batch Prediction & \$0.04 & \$0.15 \\
        Meta & Llama 3.1 (8B) & Open & Groq cloud API & \$0.05 & \$0.08 \\
        Meta & Llama 3.1 (70B) & Open & AWS Bedrock Batch Inference & \$0.36 & \$0.36 \\
        Meta & Llama 3.3 (70B) & Open & Groq Cloud API & \$0.59 & \$0.79 \\
        Mistral AI & Codestral-Mamba-2407 (7B) & Open & API & \$0.30 & \$0.90 \\
        Mistral AI & Mistral-Large-2407 (123B) & Open & API & \$2.00 & \$6.00 \\
        Mistral AI & Pixtral-Large-2411 (124B) & Open & API & \$2.00 & \$6.00 \\
        OpenAI & GPT-3.5 Turbo (175B) & Closed & Batch API & \$0.25 & \$0.75 \\
        OpenAI & GPT-4 Turbo & Closed & Batch API &\cellcolor{red!30}\$5.00 & \cellcolor{red!30}\$15.00 \\
        OpenAI & GPT-4o & Closed & Batch API & \$1.25 & \$5.00 \\
        xAI & Grok & Closed & xAI API & \$2.00 & \$10.00 \\
        \bottomrule
    \end{tabular}%
    }
\end{table}

\subsection{Code Generation}
Each LLM receives a standard prompt that consists of the problem statement, input/output specification, constraints, and example test cases (see \texttt{Appendix} for a sample prompt). To ensure {\bf fairness}, all models are tested using the exact same prompt structure. This makes sure that all performance differences stem from the model's respective behaviors rather than a deviation in the representation of the task itself. For the LLMs, the temperature parameter is set to default values (for Llama-3.1-70B it is 0.5; for Nova, Mistral, Llama-3.1-8B, and Llama-3.3-70B it is 0.7; and for the rest of the models it is 1.0).
The model will then produce an initial code solution. To ensure the {\bf correctness} of the generated code prior to measuring efficiency, each solution followed: (1) Syntax check to ensure compilability; (2) Run against 100 test cases, sourced from EffiBench; and (3) Verify against edge cases to confirm robustness.
If that solution does not compile or does not pass all 100 test cases for that problem, 
the LLM is asked to regenerate the solution. During the {\bf regeneration} phase, we provide the model with its original prompt as well as execution feedback on the code it just produced, prompting it to produce a new solution. This process can repeat up to a maximum of 25 iterations per problem, retaining the first solution that returns correct results in the interest of assessing energy and memory usage (see \texttt{Appendix}~\ref{Appendix:A} and for the details of the code generation workflow).

\subsection{Code Energy Consumption Measurements}

For collecting energy metrics, we utilize the \texttt{perf}\cite{perfstat} tool's power monitoring capabilities. Specifically, we use the 
\texttt{power/energy-pkg/} for measuring the energy consumption of the entire processor socket, including all cores and cache;
\texttt{power/energy-ram/} for measuring the energy consumption of the random access memory (RAM) attached to the integrated memory controller; and 
\texttt{cpu-clock} for measuring the execution time of the code. 
Our methodical approach consisted of several key steps to ensure the accuracy of the energy measurements: (1) Before running any problem code, we calculate the idle power consumption in the target system for a 30-second period to establish a baseline;
(2) For each execution of the problem code, we calculate the adjusted energy consumption by subtracting the baseline idle power; (3) A cooldown period of 10 seconds is implemented between executions to prevent thermal interference; and (4) Each problem code is run 5 separate times in random orders and the results are averaged to ensure statistical validity (see \texttt{Appendix}~\ref{Appendix:A.3} and \texttt{Appendix}~\ref{Appendix:B.2} for the details of the energy measurement methodology).

\subsection{Code Memory Consumption Measurements}

For collecting memory metrics, we utilize python library \texttt{Memory\_Profiler}. This library helps us sample the memory used by the process at the given intervals (in this case 0.001 seconds). 
For each problem code, we measure the \texttt{Average Memory Consumption Over Time}, which is expressed in \texttt{Megabytes * seconds} and shows how much memory a process uses and for how long, providing a cumulative view of memory consumption (see \texttt{Appendix}~\ref{Appendix:A.3} and \texttt{Appendix}~\ref{Appendix:B.3} for the details of the memory consumption measurement methodology).

\vspace{-1mm}

\subsection{Testing Environment}

To ensure fairness in our comparative analysis, all models were evaluated under identical conditions. This approach eliminated hardware and software variabilities, allowing for an effective comparison of performance metrics.
Our standardized test environment consisted of: (1) \textbf{Platform:} Chameleon Cloud \cite{chameleon_cloud} ; (2) \textbf{Processor:} Intel Xeon Gold 6126 (24 cores); (3) \textbf{Memory:} 192 GiB RAM; (4) \textbf{Operating System:} Ubuntu 24.04.1 LTS; and (5) \textbf{Kernel:} 6.8.0-51-generic.


\section{Evaluation and Analysis}

In this section, we present the findings of our experimental results. We analyze the results obtained and evaluate the performance of LLMs in terms of energy consumption and cost. We also identify the performance of LLMs compared to canonical solutions written by humans. At the same time, we present the cost breakdown of the LLMs to carry out the tasks.

\subsection{Code Generation Accuracy Analysis}

Since we do not have direct access to the server-side infrastructure of commercial LLMs to measure their computational resource usage, we develop an alternative approach to estimate the relative inference efficiency of these models during code generation. Our methodology captures both the success rate and the token-based resource consumption which serves as a proxy for computational costs.
To perform this assessment across different LLMs, we implement a systematic approach that accounts for both the success rate and the input/output tokens required for successful code generation by each model. 

Our analysis procedure follows a structured iteration approach. For each problem, we track the sequence of generation attempts until success. For each problem in our dataset (Table~\ref{table:benchmark_statistics}), we execute code generation (up to 25 times) across all LLM models under study, until a successful code passing all 100 tests is generated. This repetition allows us to calculate three key metrics:

\begin{adjustwidth}{0.6cm}{}

\vspace{-1mm}
\textbf{(1) Average Pass @}: This value indicates how many attempts are required before successfully generating a solution that passes all test cases. A lower value suggests more efficient generation.

\vspace{-1mm}
\textbf{(2) Average Total Input Tokens}: We measure the number of input prompt tokens across successful generations, which directly correlates with API costs when using LLM services.

\vspace{-1mm}
\textbf{(3) Average Total Output Tokens}: This represents the size of the generated code solutions, which also influences the operational costs of deploying these models.

\end{adjustwidth}

Table~\ref{tab:llm-performance} provides comparative performance evaluation across a variety of LLMs through Pass@1, Pass@10, and Pass@25 metrics for our entire problem set (introduced in Table~\ref{table:benchmark_statistics}). The pass metrics provide the empirical probability that at least one correct solution is generated within the first, tenth, and twenty-fifth attempt respectively across a common set of problems and provided prompt. 

\begin{table}[H]
    \caption{Comparative analysis of Large Language Models (LLMs) in terms of Pass@k accuracy and the average number of input and output tokens needed to generate the correct code.}
    \label{tab:llm-performance}
    \centering
    \scriptsize 
    \resizebox{\textwidth}{!}{%
    \begin{tabular}{l c c c c >{\centering\arraybackslash}p{1.2cm} >{\centering\arraybackslash}p{1.3cm} >{\centering\arraybackslash}p{1.8cm}}
        \toprule
         \textbf{LLM} & \textbf{Pass@1} & \textbf{Pass@10} & \textbf{Pass@25}& \textbf{Avg. Pass@} & \textbf{Avg. Input Tokens} & \textbf{Avg. Output Tokens} & \textbf{Avg. Cost of Code Generation (Cents)}\\
        \midrule
        Nova-Lite & 42.9\% & 56.6\% & 58.8\% & 2.169 & 2568.7 & 545.4 & ¢0.014 \\
        Nova-Micro & \cellcolor{red!30}33.7\% & \cellcolor{red!30}49.0\% & \cellcolor{red!30}51.4\% & 2.577 & 3875.2 & 602.7 & \cellcolor{green!40}¢0.011 \\
        Nova-Pro & 60.5\% & 73.1\% & 75.2\% & 1.896 & 1972.5 & 388.8 & ¢0.141\\
        Claude 3.5 Haiku & 70.7\% & 82.4\% & 85.4\% & 1.960 & 2108.4 & 998.7 & ¢0.568\\
        Claude 3.5 Sonnet & 77.9\% & 88.0\% & 89.7\% & 1.543 & 1598.3 & 762.6 & ¢1.623\\
        DeepSeek v3 (37B) & \cellcolor{green!40}83.6\% & \cellcolor{green!40}89.7\% & 91.1\% & 1.444 & 1411.5 & 400.6 & ¢0.163 \\
        Gemini 1.5 Flash & 65.9\% & 75.1\% & 77.3\% & 1.704 & 2122.1 & 324.7 & ¢0.013\\
        Gemini 1.5 Pro & 79.5\% & 86.9\% & 87.5\% & \cellcolor{green!40}1.255 & \cellcolor{green!40}1220.0 & 243.5 & ¢0.137\\
        Gemini 2.0 Flash & 80.5\% & 88.2\% & 89.7\% & 1.458 & 1525.9 & 364.2 & ¢0.023\\
        Gemini 2.0 Flash-Lite & 71.2\% & 81.9\% & 84.3\% & 1.734 & 1963.7 & 572.6 & ¢0.016\\
        Llama 3.1 (8B) & 53.2\% & 59.7\% & 61.8\% & 1.974 & 2251.5 & 381.0 & ¢0.014\\
        Llama 3.1 (70B) & 54.6\% & 75.2\% & 80.9\% & 2.835 & 3285.3 & 846.9 & ¢0.149 \\
        Llama 3.3 (70B) & 71.0\% & 75.2\% & 82.6\% & 3.135 & 1479.2 & 361.4 & ¢0.116\\
        Codestral-Mamba-2407 (7B) & 43.5\% & \cellcolor{red!30}44.9\% & 62.4\% & \cellcolor{red!30}7.549 & \cellcolor{red!30}11684.6 & \cellcolor{red!30}1414.1 & ¢0.478 \\
        Mistral-Large-2407 (123B) & 64.7\% & 72.7\% & 76.1\% & 1.987 & 2546.5 & 342.7 & ¢0.715 \\
        Pixtral-Large-2411 (124B) & 56.2\% & 73.5\% & 78.2\% & 2.584 & 3569.6 & 512.8 & ¢1.022\\
        GPT-3.5 Turbo (175B) & 56.8\% & 66.7\% & 70.7\% & 2.282 & 2170.7 & 401.7 & ¢0.084 \\
        GPT-4 Turbo & 56.2\% & 86.9\% & 89.3\% & 2.117 & 2258.8 & 604.4 & \cellcolor{red!30}¢2.036\\
        GPT-4o & 78.4\% & \cellcolor{green!40}89.7\% & \cellcolor{green!40}92.0\% & 1.766 & 2274.5 & 600.8 & ¢0.585\\
        Grok & 75.2\% & 83.9\% & 84.9\% & 1.415 & 1380.3 & \cellcolor{green!40}229.1 & ¢0.505 \\
        \bottomrule
    \end{tabular}%
    }
\end{table}

For all three pass rates, DeepSeek-v3, GPT-4o, Gemini 2.0 Flash, and Claude 3.5 Sonnet consistently position among the best-performing models in terms of the correctness of the code generated, demonstrating reliability with problem-solving accuracy. DeepSeek-v3 has the highest Pass@1 score of 83.6\%, whereas both DeepSeek-v3 and GPT-4o have Pass@10 scores of 89.7\% and 89.1\% respectively, and finally GPT-4o has the highest Pass@25 score of 92.0\%. Nova-Micro and Nova-Lite are the worst performers in this experiment, barely exceeding 50\% at Pass@25. 
Models like Gemini 2.0 Flash (¢0.023) and Gemini 2.0 Flash-Lite (¢0.016) demonstrate competitive Pass@10 and Pass@25 rates above 82\%, at very low costs, offering more affordable options for large-scale code generation.

\subsection{Code Energy Efficiency Analysis}

To comprehensively assess the energy efficiency of LLMs, the evaluation was conducted over two distinct benchmark sets. The first set comprises 298 common problems that all 20 LLMs successfully solved, and they formed a relatively equal distribution of algorithmic categories such as Divide \& Conquer, Binary Search, and Bit Manipulation. The second benchmark expands the analysis to a larger and more diverse set of 576 common problems that were successfully solved by 11 LLMs, providing insights into how these models handle a wider range of algorithmic complexities and problem difficulties. This provided an overall understanding of how LLMs addressed a broader class of algorithmic complexity and problem difficulty.

For each model, along with the average Pass@ rate, average input token number, and average output token number, the following key efficiency metrics were computed across all common problems:

\begin{adjustwidth}{0.6cm}{}

\vspace{-1mm}
\textbf{(1) Avg. Cost of Code Generation (Cents)}: Represents the average monetary cost required to generate a code solution for each problem using the respective LLM. 

\vspace{-1mm}
\textbf{(2) Avg. Package Energy (Joules)}: Energy consumed by the entire processor socket, including all cores and cache.

\vspace{-1mm} 
\textbf{(3) Avg. RAM Energy (Joules)}: Energy consumed by the RAM.

\vspace{-1mm}
\textbf{(4) Avg. Total Energy (Joules)}: Represents the combined energy consumption of the processor package and RAM during the execution of generated code. This metric reflects the overall energy consumption of the solution.

\vspace{-1mm}
\textbf{(5) Avg. Runtime (milliseconds)}: Captures the average time required to execute the generated solutions.

\vspace{-1mm}
 \textbf{(6) Avg. Memory Consumption (MB-sec)}: Represents the total memory usage over time during code execution, measured as the integral of memory usage over runtime. 
\vspace{-1mm}
\end{adjustwidth}

\vspace{-3mm}
\subsubsection{Benchmark Set - I: 20 LLMs \& 298 Common Problems}
Of all 20 models assessed, we recognize a subset of 298 problems from our dataset in which each LLM produces a correct answer passing all tests within the 25-regeneration limit. 
The algorithm-wise and difficulty-wise distribution of these 298 common problems is provided in Table~\ref{tab:problem-distribution-common-problems}. The majority of the problems fall under the categories of Binary Search (50), Sorting (85), and Bit Manipulation (35). In terms of difficulty, most problems are of medium complexity (179), followed by easy (89) and hard (30) problems.
This intersection guarantees that all relative efficiency experiments including energy, memory, token costs, etc. take place with a fair, unambiguous set of problems. By narrowing the evaluation emphasis to the intersection subset of problems, we remove the impact of differing model pass rates to ensure appropriate, consistent, and interpretable comparisons across models.

\begin{table}[H]
\vspace{-3mm}
    \caption{Algorithm and difficulty-wise problem distribution on Benchmark Set - I.}
    \label{tab:problem-distribution-common-problems}
    \centering
    \scriptsize
    \resizebox{\textwidth}{!}{%
    \begin{tabular}{lcccccccccccc}
        \toprule
        \textbf{Difficulty} & \textbf{Greedy} & \textbf{DP} & \textbf{Backtracking} & \textbf{Divide \& Conquer} & \textbf{DFS} & \textbf{BFS} & \textbf{Binary Search} & \textbf{Two Pointers} & \textbf{Sliding Window} & \textbf{Bit Manip.} & \textbf{Sorting} & \textbf{TOTAL} \\
        \midrule
        \textbf{Easy}   & 14 & 7  & 0  & 1 & 3 & 0  & 13 & 22 & 4  & 18  & 33 & \cellcolor{gray!25}89  \\
        \textbf{Medium} & 46 & 49 & 10 & 3 & 15 & 13 & 21 & 28 & 17 & 14 & 51 & \cellcolor{gray!25}179 \\
        \textbf{Hard}   & 6  & 18 & 1  & 0 & 2  & 3  & 4  & 0  & 2  & 3  & 1  & \cellcolor{gray!25}30  \\
        \textbf{TOTAL}  & \cellcolor{gray!25}66 & \cellcolor{gray!25}74 & \cellcolor{gray!25}11 & \cellcolor{gray!25}4 & \cellcolor{gray!25}20 & \cellcolor{gray!25}16 & \cellcolor{gray!25}38 & \cellcolor{gray!25}50 & \cellcolor{gray!25}23 & \cellcolor{gray!25}35 & \cellcolor{gray!25}85 & \cellcolor{gray!25}298 \\
        \bottomrule
    \end{tabular}%
    }
\end{table}


 We present the evaluation results for this subset of problems in this section.
The performance and resource utilization of human-written canonical solutions, compared to the LLMs, are illustrated in Table~\ref{tab:llm-all-results}.  Despite recent advancements in model correctness, as measured by Pass@ rates, the canonical solutions consistently outperform all evaluated LLMs across key sustainability metrics, including energy consumption and memory efficiency.

\textbf{Average Total Energy Consumption:} On average, canonical solutions require only 5.77~J, significantly lower than any LLM-generated solutions. The most efficient LLM, DeepSeek-v3, consumes 5.91~J, while others range from 6.12~J to 12.00~J.

\textbf{Average Runtime:} Human-written code executes in just 74.16~ms, outperforming all LLM-generated solutions, which exhibit runtimes ranging from 75.64~ms (DeepSeek-v3) to 147.95~ms (GPT-4 Turbo).

\textbf{Average Memory Usage:} Canonical solutions also demonstrate superior memory efficiency with an average memory consumption of 8.70~MB$\cdot$s, lower than nearly all LLM-generated solutions. Only DeepSeek-v3 (8.67~MB$\cdot$s) achieves comparable memory efficiency, while other LLM-generated solutions exhibit higher memory consumption, ranging from 9.02~MB$\cdot$s to 11.57~MB$\cdot$s.

\begin{table}[H]
    \caption{Performance and resource usage comparison of LLMs against canonical solutions on Benchmark Set - I.}
    \label{tab:llm-all-results}
    \centering
    \scriptsize
    \resizebox{\textwidth}{!}{%
    \begin{tabular}{l >{\centering\arraybackslash}p{1.8cm} >{\centering\arraybackslash}p{1.2cm} 
                    >{\centering\arraybackslash}p{1.3cm} >{\centering\arraybackslash}p{1cm} 
                    >{\centering\arraybackslash}p{1.1cm} >{\centering\arraybackslash}p{1.1cm} 
                    >{\centering\arraybackslash}p{1.1cm} >{\centering\arraybackslash}p{1.3cm} 
                    >{\centering\arraybackslash}p{1.1cm}}
        \toprule
        \textbf{Model} & \textbf{Avg. Cost of Code Generation (Cents)} & \textbf{Avg. Input Tokens} 
        & \textbf{Avg. Output Tokens} & \textbf{Avg. Pass@} & \textbf{Avg. Pkg Energy (J)} 
        & \textbf{Avg. RAM Energy (J)} & \textbf{Avg. Total Energy (J)} 
        & \textbf{Avg. Runtime (ms)} & \textbf{Avg. Mem (MB·s)} \\
        \midrule
        \textbf{Canonical Solution} & \cellcolor{gray!25}-- & \cellcolor{gray!25}-- & \cellcolor{gray!25}-- 
        & \cellcolor{gray!25}-- & \cellcolor{gray!25}5.05 & \cellcolor{gray!25}0.72 
        & \cellcolor{gray!25}5.77 & \cellcolor{gray!25}74.16 & \cellcolor{gray!25}8.70 \\

        DeepSeek v3 & ¢0.084 & 789.6 & 146.0 & 1.017 & \cellcolor{green!40}5.17 
        & \cellcolor{green!40}0.74 & \cellcolor{green!40}5.91 & \cellcolor{green!40}75.64 & \cellcolor{green!40}8.67 \\

        Gemini 2.0 Flash & ¢0.012 & 872.4 & 166.9 & 1.020 & 5.36 & 0.76 & 6.12 & 78.05 & 9.02 \\

        Claude 3.5 Sonnet & \cellcolor{red!30}¢1.029 & 886.5 & 508.4 & 1.047 & 5.61 & 0.80 & 6.41 
        & 81.21 & 9.04 \\

        GPT-4o & ¢0.179 & \cellcolor{green!40}784.0 & 161.2 & 1.010 & 5.93 & 0.84 & 6.77 
        & 85.54 & 9.32 \\

        Nova-Lite & ¢0.007 & 1303.2 & 267.5 & 1.409 & 6.07 & 0.86 & 6.93 & 87.55 & 9.26 \\

        Claude 3.5 Haiku & ¢0.323 & 1049.1 & 597.5 & 1.191 & 6.22 & 0.88 & 7.10 & 89.97 & 9.46 \\

        Nova-Pro & ¢0.080 & 1147.3 & 214.7 & 1.272 & 6.24 & 0.88 & 7.12 & 90.03 & 9.58 \\

        Gemini 2.0 Flash-Lite & \cellcolor{green!40}¢0.006 & 934.9 & 181.2 & 1.091 & 6.26 & 0.88 
        & 7.14 & 90.44 & 9.56 \\

        GPT-3.5 Turbo & ¢0.042 & 1129.9 & 182.3 & 1.339 & 6.30 & 0.89 & 7.19 & 91.17 & 9.66 \\

        Pixtral-Large-2411 & ¢0.361 & 1288.0 & 173.0 & 1.188 & 6.55 & 0.91 & 7.46 & 94.19 & 9.18 \\

        Codestral-Mamba-2407 & ¢0.279 & \cellcolor{red!30}6973.3 & \cellcolor{red!30}773.2 
        & \cellcolor{red!30}4.379 & 6.90 & 0.96 & 7.86 & 99.19 & 9.44 \\

        Llama 3.1 (8B) & ¢0.070 & 943.7 & 178.8 & 1.581 & 7.11 & 0.99 & 8.10 & 101.88 & 9.72 \\

        Nova-Micro & ¢0.007 & 2132.8 & 438.5 & 2.067 & 7.12 & 0.99 & 8.11 & 102.28 & 10.21 \\

        Gemini 1.5 Pro & ¢0.089 & 835.1 & 146.8 & \cellcolor{green!30}1.007 & 7.60 & 1.06 & 8.66 
        & 108.32 & 9.65 \\

        Gemini 1.5 Flash & \cellcolor{green!40}¢0.006 & 1007.3 & 163.3 & 1.074 & 8.38 & 1.16 & 9.53 
        & 119.33 & 10.30 \\

        Llama 3.3 (70B) & ¢0.054 & 1257.2 & 249.9 & 1.342 & 8.60 & 1.19 & 9.79 & 122.35 & 10.85 \\

        Mistral-Large-2407 & ¢0.362 & 1274.6 & 177.6 & 1.228 & 8.63 & 1.19 & 9.82 & 122.42 & 10.49 \\

        Grok 2 & ¢0.281 & 799.9 & \cellcolor{green!40}121.2 & 1.013 & 8.77 & 1.21 & 9.98 
        & 124.50 & 10.11 \\

        Llama 3.1 (70B) & \cellcolor{green!40}¢0.006 & 977.0 & 159.6 & 1.185 & 8.87 & 1.23 & 10.10 
        & 126.07 & 10.39 \\

        GPT-4 Turbo & ¢0.729 & 881.9 & 192.3 & 1.091 & \cellcolor{red!30}10.6 & \cellcolor{red!30}1.44 
        & \cellcolor{red!30}12.00 & \cellcolor{red!30}147.95 & \cellcolor{red!30}11.57 \\
        \bottomrule
    \end{tabular}%
    }
\end{table}

When we analyze the results based on the problem difficulty level, we observe that all LLMs perform similarly to canonical solutions on easy problems. For medium-difficulty problems, solutions generated by DeepSeek-v3 consume approximately 1.02 times the energy of canonical solutions, while Nova-Lite consumes around 1.20 times. GPT-4 Turbo performes the worst on medium problems, consuming 2.08 times more energy than canonical solutions. For hard problems, DeepSeek-v3 performs comparably to canonical solutions in terms of energy efficiency, while solutions generated by Llama-3.1-70B consume approximately 1.40 times more energy.

Analyzing performance across different algorithmic categories, LLM-generated solutions for problems involving BFS, DFS, and Two-Pointer algorithms achieve energy efficiency similar to canonical solutions. However, LLM-generated solutions for problems involving Sorting and Dynamic Programming consistently require more energy. For Binary Search and Bit Manipulation problems, most LLMs generate code that is as efficient as canonical solutions, except for Llama-3.1-70B and Llama-3.3-70B, which produce solutions consuming significantly more energy. Across all algorithms and difficulty levels, DeepSeek-v3 and GPT-4o consistently outperform other LLMs in terms of energy efficiency and runtime performance (see \texttt{Appendix}~\ref{Appendix:D} and \texttt{Appendix}~\ref{Appendix:E}
for the detailed results based on the problem difficulty level and across different algorithmic categories).

\subsubsection{Benchmark Set - II: 11 LLMs \& 576 Common Problems}

Table~\ref{tab:problem-distribution-batch2} presents the algorithm-wise and difficulty-level breakdown of the 576 problems included in Benchmark Set II. These 576 problems were successfully passed by 11 LLMs within the 25-regeneration limit. A significant portion of these problems fall under the categories of Dynamic Programming (158), Greedy algorithms (149), and Sorting (155). With respect to difficulty, the majority of problems are of medium complexity (347), followed by hard (110) and easy (119).

\begin{table}[H]
\vspace{-3mm}
    \caption{Algorithm and difficulty-wise problem distribution on Benchmark Set - II.}
    \label{tab:problem-distribution-batch2}
    \centering
    \scriptsize
    \resizebox{\textwidth}{!}{%
    \begin{tabular}{lcccccccccccc}
        \toprule
        \textbf{Difficulty} & \textbf{Greedy} & \textbf{DP} & \textbf{Backtracking} & \textbf{Divide \& Conquer} & \textbf{DFS} & \textbf{BFS} & \textbf{Binary Search} & \textbf{Two Pointers} & \textbf{Sliding Window} & \textbf{Bit Manip.} & \textbf{Sorting} & \textbf{TOTAL} \\
        \midrule
        \textbf{Easy}   & 28  & 7   & 1  & 3  & 3  & 0  & 16  & 24  & 7  & 22  & 48  & \cellcolor{gray!25}119 \\
        \textbf{Medium} & 101 & 98  & 19 & 4  & 27 & 24 & 46  & 37  & 31 & 34  & 89  & \cellcolor{gray!25}347 \\
        \textbf{Hard}   & 20  & 53  & 5  & 5  & 3  & 6  & 24  & 1   & 10 & 10  & 18  & \cellcolor{gray!25}110 \\
        \textbf{TOTAL}  & \cellcolor{gray!25}149 & \cellcolor{gray!25}158 & \cellcolor{gray!25}25 & \cellcolor{gray!25}12 
                         & \cellcolor{gray!25}33  & \cellcolor{gray!25}30  & \cellcolor{gray!25}86 & \cellcolor{gray!25}62 
                         & \cellcolor{gray!25}48  & \cellcolor{gray!25}66  & \cellcolor{gray!25}155 & \cellcolor{gray!25}576 \\
        \bottomrule
    \end{tabular}%
    }
\end{table}

Table~\ref{tab:llm-resources-final} summarizes the performance and resource costs of human-written canonical solutions compared to LLM-generated solutions for the second benchmark set. The problems in this set are more complex, and the canonical solutions consistently outperform the LLM-generated solutions across all relevant efficiency metrics. Moreover, while some LLMs demonstrate better model correctness through higher Pass@ rates, these improvements are often accompanied by significant increases in energy and memory consumption.

\textbf{Average Total Energy Consumption:} On average, canonical solutions require only 5.46~J, substantially lower than any LLM-generated solutions. While DeepSeek-v3 remains the most energy-efficient LLM (6.37~J), it still consumes 16.7\% more energy than the canonical solutions. Interestingly, Gemini-1.5-Pro exhibits the best Avg. Pass@ score (1.056) for this set of problems, but its energy consumption is among the highest at 11.15~J.

\textbf{Average Runtime:} Human-written code maintains the lowest average runtime at 69.36~ms, consistently outperforming all LLM-generated solutions. Although smaller models typically demonstrate faster runtimes, some models like Claude-3.5-Haiku still require 88.56~ms. Larger models such as Llama-3.1-70B and Llama-3.3-70B exhibit even higher runtimes at 112.01~ms and 130.02~ms, respectively. The highest runtime is by Gemini 1.5 Pro at 137.14~ms.

\begin{table}[H]
    \caption{Performance and resource usage comparison of LLMs against canonical solutions on Benchmark Set - II.}
    \label{tab:llm-resources-final}
    \centering
    \scriptsize
    \resizebox{\textwidth}{!}{%
    \begin{tabular}{l >{\centering\arraybackslash}p{1.8cm} >{\centering\arraybackslash}p{1.5cm} 
                    >{\centering\arraybackslash}p{1.5cm} >{\centering\arraybackslash}p{1cm} 
                    >{\centering\arraybackslash}p{1.5cm} >{\centering\arraybackslash}p{1.5cm} 
                    >{\centering\arraybackslash}p{1.5cm} >{\centering\arraybackslash}p{1.2cm} 
                    >{\centering\arraybackslash}p{1.5cm}}
        \toprule
        \textbf{Model} & \textbf{Avg. Cost of Code Generation (Cents)} & \textbf{Avg. Input Tokens} 
        & \textbf{Avg. Output Tokens} & \textbf{Avg. Pass@} & \textbf{Avg. Pkg Energy (J)} 
        & \textbf{Avg. RAM Energy (J)} & \textbf{Avg. Total Energy (J)} & \textbf{Avg. Runtime (ms)} 
        & \textbf{Avg. Mem (MB·s)} \\
        \midrule
        \textbf{Canonical Solution} & \cellcolor{gray!25}-- & \cellcolor{gray!25}-- & \cellcolor{gray!25}-- 
        & \cellcolor{gray!25}-- & \cellcolor{gray!25}4.78 & \cellcolor{gray!25}0.68 
        & \cellcolor{gray!25}5.46 & \cellcolor{gray!25}69.36 & \cellcolor{gray!25}6.62 \\

        DeepSeek v3 & ¢0.097 & \cellcolor{green!40}889.5 & 191.0 & 1.075 & \cellcolor{green!40}5.59 
        & \cellcolor{green!40}0.78 & \cellcolor{green!40}6.37 & \cellcolor{green!40}80.45 & \cellcolor{green!40}6.90 \\

        GPT-4o & ¢0.226 & 934.6 & 217.7 & 1.092 & 5.82 & 0.81 & 6.63 & 83.54 & 7.44 \\

        Claude 3.5 Sonnet & ¢1.088 & 964.0 & 532.3 & 1.076 & 5.91 & 0.82 & 6.74 & 84.60 & 7.33 \\

        Claude 3.5 Haiku & ¢0.401 & 1420.8 & \cellcolor{red!30}718.7 & 1.415 & 6.19 & 0.86 & 7.05 
        & 88.56 & 7.65 \\

        Gemini 2.0 Flash-Lite & \cellcolor{green!40}¢0.009 & 1230.1 & 241.4 & 1.307 & 6.42 & 0.89 
        & 7.31 & 91.91 & 8.27 \\

        Llama 3.1 (70B) & ¢0.116 & \cellcolor{red!30}2562.4 & 644.6 & 2.309 & 7.91 & 1.09 & 9.00 
        & 112.01 & 8.22 \\

        GPT-4 Turbo & \cellcolor{red!30}¢1.204 & 1377.8 & 343.4 & 1.448 & 8.27 & 1.13 & 9.40 
        & 117.10 & 9.08 \\

        Gemini 2.0 Flash & ¢0.016 & 1084.4 & 231.5 & 1.151 & 8.63 & 1.17 & 9.80 & 121.34 & 7.77 \\

        Llama 3.3 (70B) & ¢0.097 & 1248.6 & 291.4 & \cellcolor{red!30}2.439 & 9.26 & 1.26 & 10.52 
        & 130.02 & \cellcolor{red!30}10.73 \\

        Grok 2 & ¢0.352 & 981.8 & \cellcolor{green!40}156.0 & 1.146 & 9.62 & 1.31 & 10.93 
        & 135.28 & 9.34 \\

        Gemini 1.5 Pro & ¢0.102 & 927.2 & 174.1 & \cellcolor{green!40}1.056 & \cellcolor{red!30}9.82 
        & \cellcolor{red!30}1.33 & \cellcolor{red!30}11.15 & \cellcolor{red!30}137.14 & 8.43 \\
        \bottomrule
    \end{tabular}%
    }
\end{table}

\textbf{Average Memory Usage:} Canonical solutions also demonstrate superior memory efficiency with an average memory consumption of 6.62~MB$\cdot$s. These values are consistently lower than those of any LLM-generated solutions in the study, where memory usage ranges from 6.90~MB$\cdot$s (DeepSeek-v3) to a maximum of 10.73~MB$\cdot$s (Llama-3.3-70B).

When we analyze the results based on difficulty, for easy problems all LLM-generated code performs similarly to that of canonical solutions. Llama-3.1-70B and Llama-3.3-70B models are comparatively better in giving energy-efficient solutions to medium problems than hard problems. Analyzing the results for various algorithms, DeepSeek-v3 consistently performs better than all other LLMs in almost all categories, except for BFS, Backtracking, and Bit Manipulation. Llama models (3.1-70B, 3.3-70B) perform poorly across almost all algorithmic categories. While Grok-2 performs relatively worse than all other LLMs for Dynamic Programming and Binary Search, it gives more energy-efficient solutions to problems involving DFS, Two Pointers, Sorting, and Greedy algorithms. Gemini-1.5 Pro performs poorly in Dynamic Programming and Bit Manipulation while performing relatively well in other categories. GPT-4 Turbo performs poorly in Sorting-based problems while doing better in others.



\section{Discussion \& Conclusions}

This research provides a systematic, empirical approach to comparing energy consumption, runtime, memory usage, and inference costs of code generated by state-of-the-art LLMs and human-generated solutions. 878 coding problems from the EffiBench framework were evaluated, and while they show that LLMs achieve high Pass@N rates and functional correctness, their energy efficiency is always poorer.
Our results highlight that LLMs are inefficient mainly due to duplicate logic, inappropriate decisions of algorithms, and excessive memory requirements for trivial tasks.
While both DeepSeek-v3 and GPT-4o perform the best across all LLMs used in this study, they still require more energy and memory than human-generated code.

Our comprehensive study shows that LLMs can generate functionally correct code, but they may incur significantly large energy consumption.
The concerns for their resource efficiency are influenced by earlier evaluations with token-based energy metrics with Pass@N scores which demonstrated that time taken in latency can be higher for correctness (in terms of larger LLMs), not only due to increased complexity of input, increased code length, and having to repeat many iterations to fix syntactic bugs.

We demonstrate that LLM performance is highly dependent on problem complexity and algorithm type. In Benchmark Set-I, LLMs perform well and produce solutions close to canonical implementations due to the lower problem complexity which is mostly on simpler categories such as Binary Search and Divide and Conquer. However, in Benchmark Set-II, with a higher proportion of complex problems involving Dynamic Programming, Greedy algorithms, and Sorting, the inefficiencies of LLMs become more apparent, which shows their limitations in handling computationally demanding tasks.

To conclude, sustainable software generation requires balancing correctness with computational efficiency. As LLMs become integrated into software development, it is critical to adopt evaluation frameworks that account for energy and memory usage to ensure both functional and environmental viability.

{\bf Limitations:}
Though this work performs a comprehensive study in evaluating LLMs' ability to generate energy-efficient code, it has some limitations: {\bf (1)} LLMs are not deterministic. The same LLM might generate a different code, the next time we ask it. {\bf (2)} Datasets used could have been seen by some LLMs during the training phase, resulting in it memorizing and generating the most efficient solution. {\bf (3)} This work is currently limited to evaluating codes generated in Python only and does not consider other programming languages. 

{\bf Future Work:} To address the above limitations, future studies can focus on including a diverse range of programming languages and problems that were created recently but not seen by LLMs during their training phase. The impact of various prompt engineering techniques on the code generation aspect of LLMs can also be thoroughly investigated. 

\section* {Acknowledgments}
This project is in part sponsored by the National Science Foundation (NSF) under award number CBET-2343284, by the State University of New York (SUNY), and by IBM Corporation. We also thank Chameleon Cloud for making their resources available for the experiments of this work. 

\small
\bibliographystyle{unsrt}
\bibliography{references}






\newpage
\section*{APPENDIX}
\appendix



\section{Experimentation Details}
\label{Appendix:A}

\subsection{End-to-End Experimentation Workflow}
\label{Appendix:A.1}

\begin{figure}[H]
    \centering
    \includegraphics[width=0.8\linewidth]{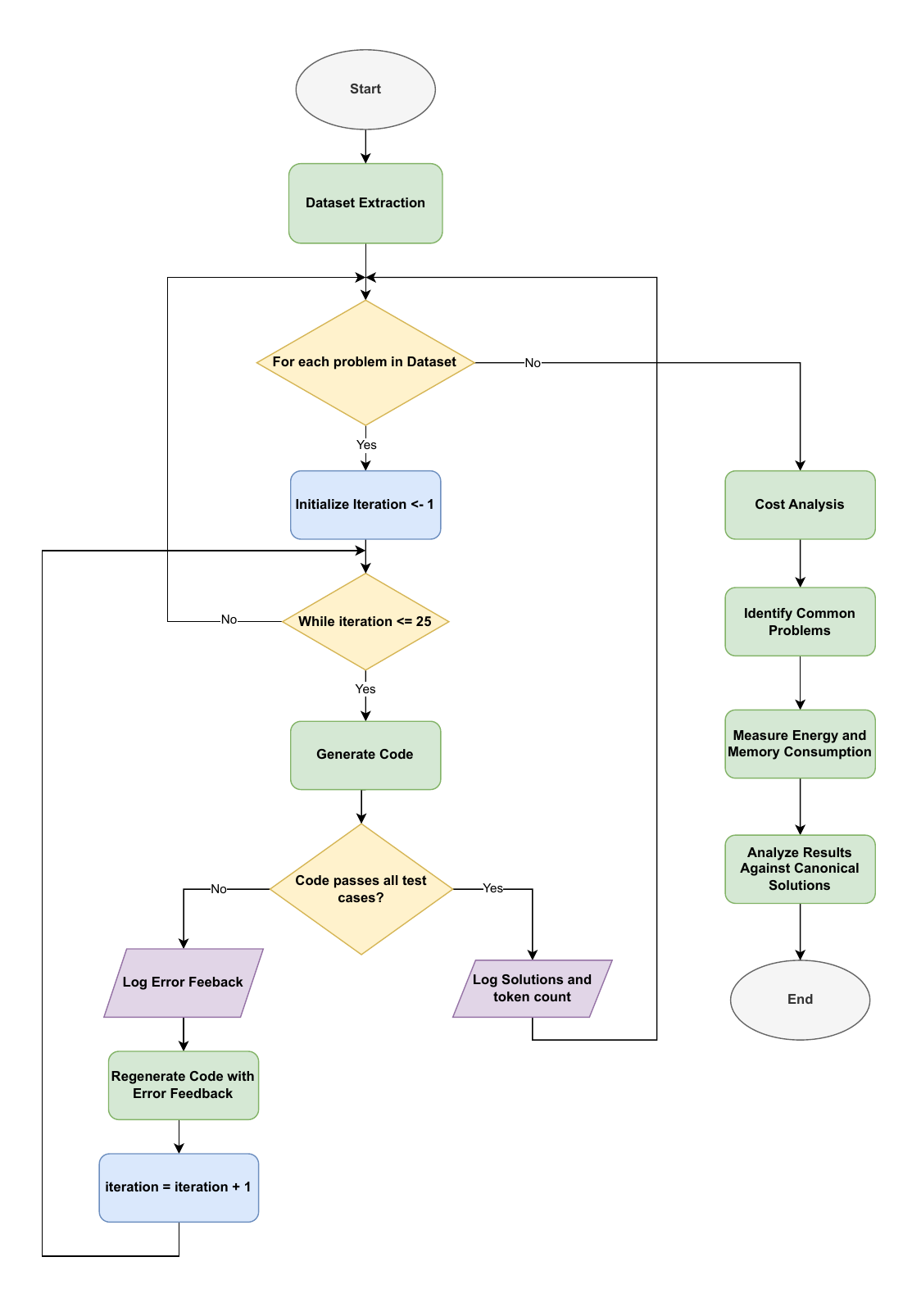}
    \caption{End-to-end experimentation workflow.}
    \label{fig:overall_workflow}
\end{figure}

The end-to-end experimentation workflow of this study is shown in Figure. \ref{fig:overall_workflow}. We consider 878 problems from \textsc{EffiBench} (detailed in Table~\ref{table:benchmark_statistics}) and generate solutions using 20 popular LLMs (discussed in Table \ref{tab:llm-table}). 

The source code of the end-to-end experimentation workflow is made available to the public and it can be found at the following repository:  \url{https://anonymous.4open.science/r/evaluating-the-energy-efficiency-of-the-code-generated-by-llms-0D72}.

\subsection{LLM Code Generation}
\label{Appendix:A.2}
\begin{figure}[H]
    \centering
    \includegraphics[width=0.8\linewidth]{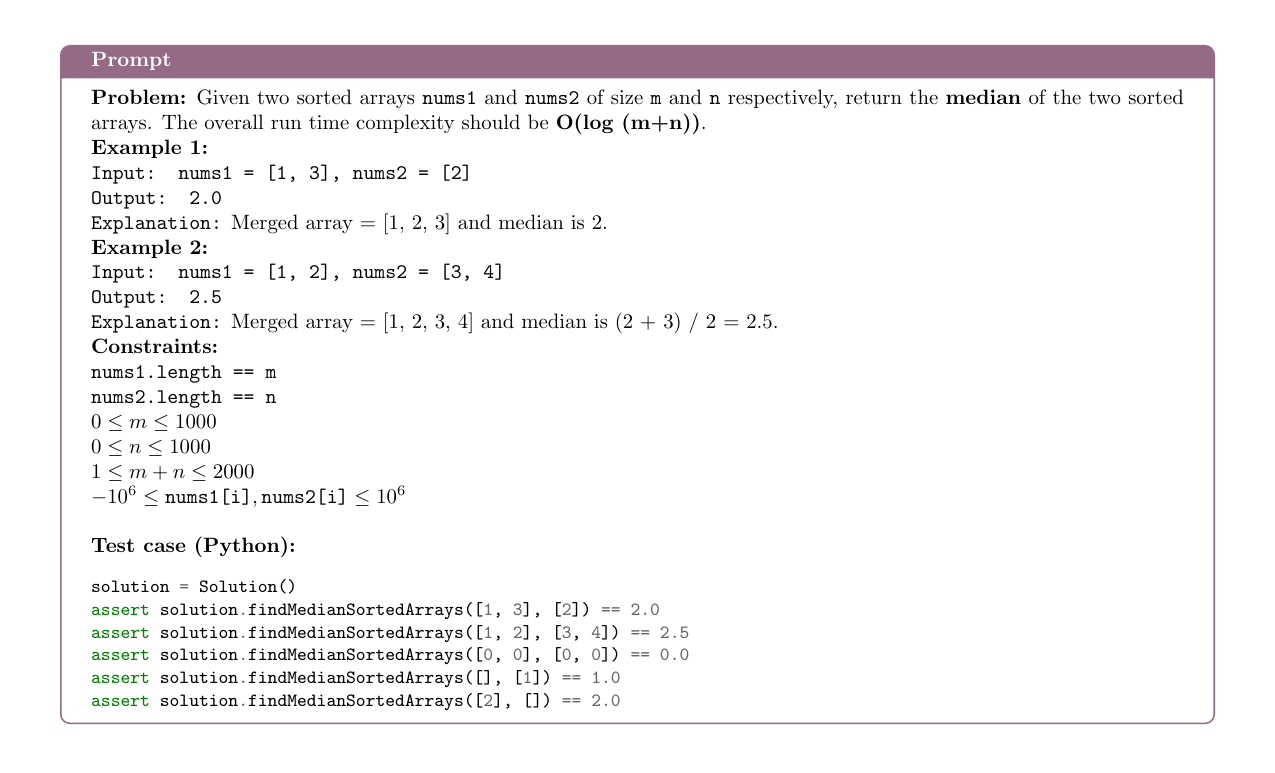}
    \caption{Prompt structure used for code generation.}
    \label{fig:prompt}
\end{figure}

We use a prompt structure similar to the one used in \textsc{Effibench}, which follows the MBPP code generation prompt, as shown in Figure \ref{fig:prompt}. To ensure fairness, we use the same prompt for all LLMs.

\begin{figure}[H]
    \centering
    \includegraphics[width=0.7\linewidth]{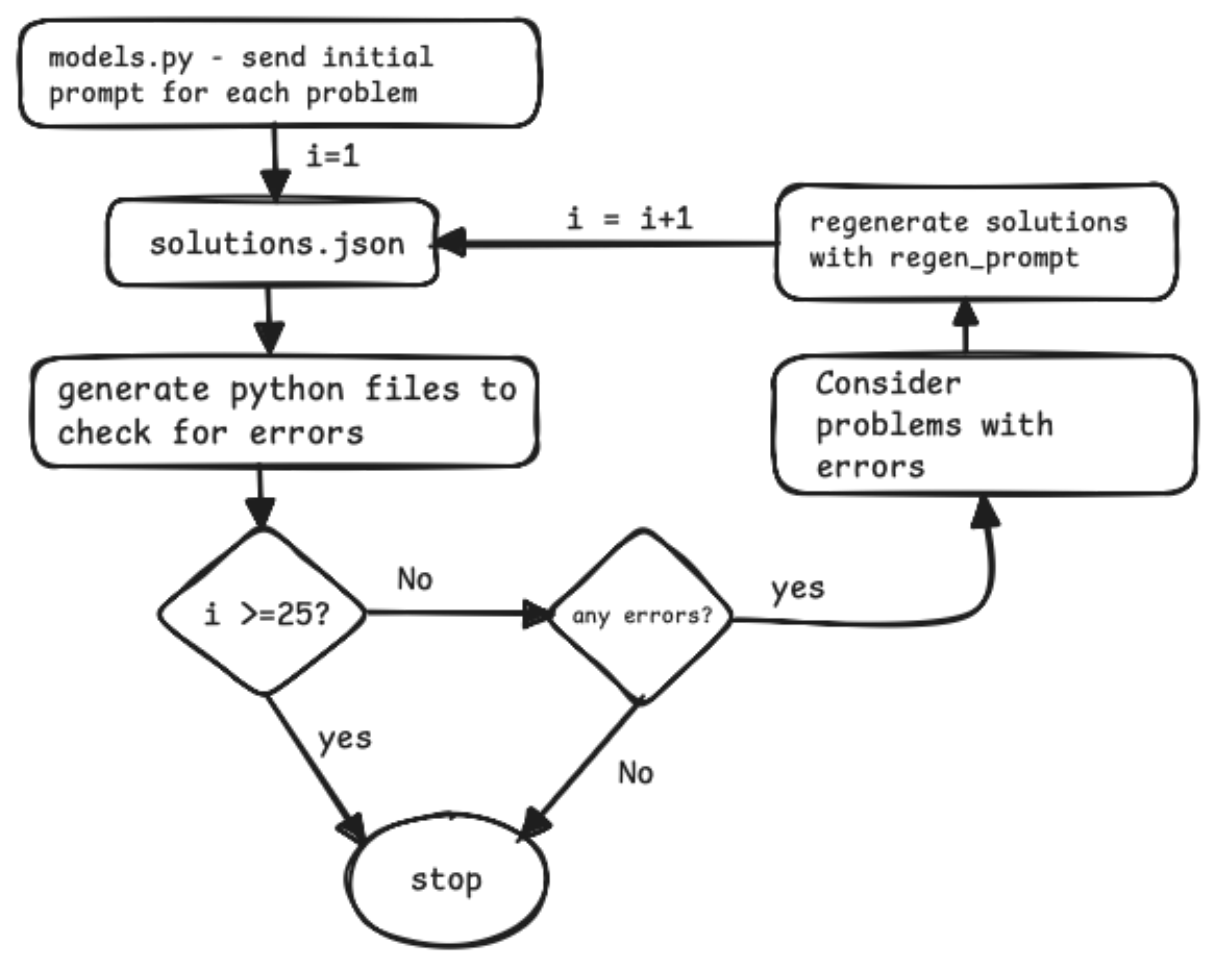}
    \caption{The methodology followed for code generation.}
    \label{fig:code_gen_workflow}
\end{figure}

The methodology followed for generating code with LLMs is shown in Figure \ref{fig:code_gen_workflow}. We use batch inference wherever possible to save on time and costs associated with code generation. If the LLMs are not able to generate the correct code in the first go, retries are allowed till it generates a correct solution or up to 24 times. After each iteration, python files for each problem are generated, they contain the solution generated by LLMs followed by test cases in the dataset. All the files are then executed and errors are logged. In the next regeneration, the error message is included in the prompt to correct its mistake. To ensure fairness, all models follow the same prompt format. Once a correct solution is generated, that problem will not be considered for regeneration in the next iteration. During each iteration, the number of input tokens and the number of output tokens to LLMs are noted along with the number of problems passed till that iteration. For code regeneration, along with the initial prompt, we also pass the incorrect solution given by LLM, along with the error message, to help it correct the result in the next output.

\subsection{Energy, Memory, and Runtime Measurements}
\label{Appendix:A.3}

\begin{figure}[H]
    \centering
    \includegraphics[width=0.8\linewidth]{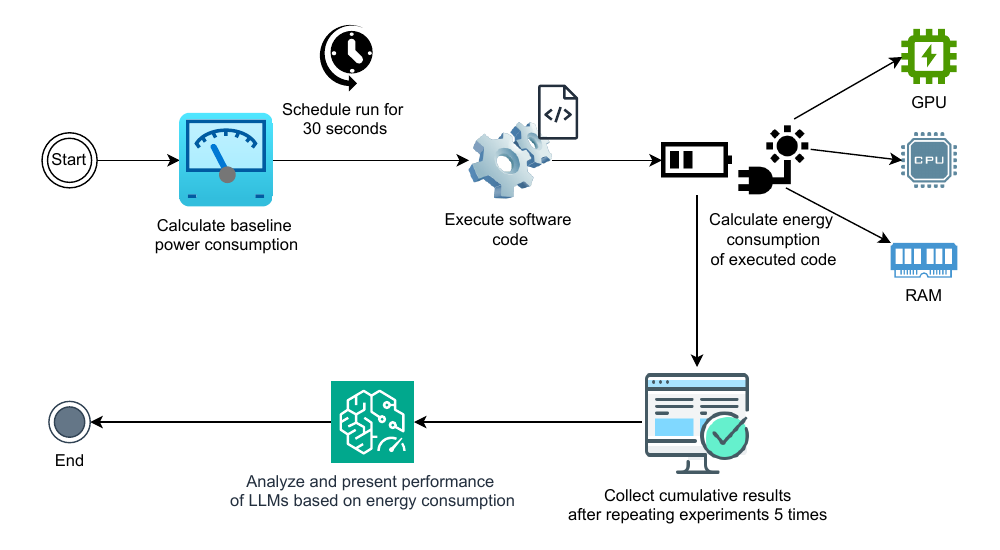}
    \caption{{Flowchart showing the experimental process.}}
    \label{fig:experimental}
\end{figure}

{Figure \ref{fig:experimental} shows the experimentation process we followed for measuring the runtime, memory, and energy consumed by each tested code.}

For collecting energy metrics, we utilize the \texttt{perf} tool's power monitoring capabilities, specifically the \texttt{power/energy-pkg/}, \texttt{power/energy-ram/}, and \texttt{cpu-clock} packages.

We measure the baseline power consumption of the system while no process is running on it for 30 seconds.
\begin{equation}
P_{\text{baseline}} = \frac{E_{\text{idle, 30s}}}{30\,\text{s}}
\end{equation}
Then, we calculate the energy consumed by code $E_{\text{adjusted}}$ as,
\begin{equation}
\label{eq:adj_energy}
   E_{\text{adjusted}} = E_{\text{actual}} - P_{\text{baseline}} \times t_{\text{code}}
\end{equation}

where $E_{\text{actual}}$ is the energy reported by Perf while the code runs for $t_{\text{code}}$ seconds.

Baseline power consumption is measured for 30 seconds before processing each code, and a cool-down period of 10 seconds is maintained between each run to reduce the effect of previous executions. Each problem code is run 5 separate times in random orders and the results are averaged to ensure statistical validity. 

In addition to the above experiment, we calculate the memory used by each solution using the python library \texttt{Memory\_Profiler}. This library helps us sample the memory used by the process at the given intervals (in this case 0.001 seconds) and store the readings in a .dat file for each solution. This experiment is done 3 times independently of the previous one with perf and the results are averaged to reduce noise.

The measured results are analyzed using the metrics discussed in Appendix \ref{Appendix:B}.

\section{Evaluation Metrics and Measurements} \label{Appendix:B}

\subsection{LLM Inference Cost Analysis}
\label{Appendix:B.1}

For each LLM model $M$ and problem $p$, we define the following metrics:

\begin{align}
\text{Pass@}(M, p) &= \min\{i \in \{1, 2, \ldots, 25\} \mid \text{solution at iteration } i \text{ passes all test cases}\} \\
\text{InputTokens}(M, p) &= \sum_{i=1}^{\text{Pass@}(M, p)} \text{InputTokens}_i(M, p) \\
\text{OutputTokens}(M, p) &= \sum_{i=1}^{\text{Pass@}(M, p)} \text{OutputTokens}_i(M, p)
\end{align}

where $\text{InputTokens}_i(M, p)$ represents the number of input tokens for iteration $i$ and $\text{OutputTokens}_i(M, p)$ represents the number of output tokens for iteration $i$.

The aggregate metrics across all problems $P$ for a model $M$ are computed as:

\begin{align}
\text{Avg. Pass@}(M) &= \frac{1}{|P|}\sum_{p \in P} \text{Pass@}(M, p) \\
\text{Avg. InputTokens}(M) &= \frac{1}{|P|}\sum_{p \in P} \text{InputTokens}(M, p) \\
\text{Avg. OutputTokens}(M) &= \frac{1}{|P|}\sum_{p \in P} \text{OutputTokens}(M, p)
\end{align}


\subsection{Code Energy Consumption Metrics}
\label{Appendix:B.2}

\textbf{Execution Time:} represents the time taken by the problem to complete execution, represented by $T_{code}$. It is measured using \textit{perf}.

\textbf{Average Execution Time:} is the sum of the execution times of problems considered for the experiment divided by the number of problems considered (N).
\begin{equation}
\text{Avg. ET} = \frac{1}{N} \sum_{i=1}^{N} T_{\text{code}_i}
\end{equation}

\textbf{Package Energy:} is the energy consumed by the entire processor socket, including all cores and cache, during the execution of code as measured by perf. It is the adjusted package energy calculated using equation \ref{eq:adj_energy}.

\textbf{RAM Energy:} is the energy consumed by random access memory (RAM) during the execution of the code, measured using perf. It is then adjusted considering the baseline power consumption using the equation \ref{eq:adj_energy}.

\textbf{Total Energy:} is the sum of package energy and RAM energy for a particular problem.

\textbf{Average Total Energy:} is the sum of the total energy consumed by problems considered for the experiment divided by the number of problems considered.

\begin{equation}
\text{Avg. TE} = \frac{1}{N} \sum_{i=1}^{N} TE_{\text{code}_i}
\end{equation}

\subsection{Code Memory Consumption Metrics}
\label{Appendix:B.3}

{\bf Memory Consumption Over Time:}

This metric measures how much memory a process uses and for how long, providing a cumulative view of memory consumption:

\begin{equation}
\text{mem-sec} = \sum_{i=1}^{n-1} \left(\frac{M_i + M_{i+1}}{2} \times (T_{i+1} - T_i)\right)
\end{equation}

Where:
\begin{itemize}
    \item $M_i$ is the memory consumption at time $T_i$
    \item $n$ is the number of sampling points
    \item The unit is memory-seconds (MB-sec)
\end{itemize}





The final memory metric for a model $M$ and problem $p$ is derived as:

\begin{equation}
\text{FinalMemory}(M, p) = \frac{1}{3} \sum_{i=1}^{3} \text{Memory}_{i}(M, p)
\end{equation}

where $\text{Memory}_{i}(M, p)$ is the memory measurement (either \texttt{mem-sec}) from the $i^{\text{th}}$ run.

\section{Comparative Relative Efficiency Analysis}
\label{Appendix:C}

For each LLM-generated code solution and a given resource metric $r_i$, the relative cost quantifies how many times more or fewer resources the LLM-generated output used compared to the canonical solution. 

\begingroup
\vspace{-3mm}
\begin{equation}
\text{Relative}_{m, r_i} = \frac{V_{m, r_i}}{V_{\text{canonical}, r_i}}
\end{equation}
\vspace{-2mm}
\endgroup

\noindent
Where:
\begin{itemize}[noitemsep,nolistsep]
    \item $V_{m, r_i}$: value of model $m$ for metric $r_i$
    \item $V_{\text{canonical}, r_i}$: value of the canonical solution for the same metric
    \item $\text{Relative}_{m, r_i}$: relative cost of the model $m$ compared to the canonical solution
\end{itemize}

\noindent
A value greater than 1 indicates that the LLM used more resources (i.e., was less efficient) than the canonical solution. A value less than 1 indicates that the LLM used fewer resources. Values close to 1 reflect near-equivalent cost.



\begin{table}[H]
    \caption{Relative cost of LLMs compared to canonical solutions on \textbf{Benchmark Set - I}.}
    \label{tab:relative-efficiency-b1}
    \centering
    \scriptsize
    \resizebox{\textwidth}{!}{%
    \begin{tabular}{lccccc}
        \toprule
        \textbf{Model} & \textbf{Avg. Package Energy} & \textbf{Avg. RAM Energy} & \textbf{Avg. Total Energy} & \textbf{Avg. Runtime} & \textbf{Avg. Memory} \\
        \midrule
        DeepSeek v3 & 1.0238 & 1.0278 & 1.0243 & 1.0200 & 0.9966 \\
        Gemini 2.0 Flash & 1.0614 & 1.0556 & 1.0607 & 1.0525 & 1.0368 \\
        Claude 3.5 Sonnet & 1.1109 & 1.1111 & 1.1109 & 1.0951 & 1.0391 \\
        GPT-4o & 1.1743 & 1.1667 & 1.1733 & 1.1535 & 1.0713 \\
        Nova-Lite & 1.2020 & 1.1944 & 1.2010 & 1.1806 & 1.0644 \\
        Claude 3.5 Haiku & 1.2307 & 1.2222 & 1.2296 & 1.2131 & 1.0874 \\
        Nova-Pro & 1.2356 & 1.2222 & 1.2345 & 1.2132 & 1.1011 \\
        Gemini 2.0 Flash-Lite & 1.2396 & 1.2222 & 1.2385 & 1.2195 & 1.0985 \\
        GPT-3.5 Turbo & 1.2475 & 1.2361 & 1.2470 & 1.2389 & 1.1118 \\
        Pixtral-Large-2411 & 1.2970 & 1.2639 & 1.2938 & 1.2704 & 1.0552 \\
        Codestral-Mamba-2407 & 1.3663 & 1.3333 & 1.3629 & 1.3373 & 1.0851 \\
        Llama 3.1 (8B) & 1.4079 & 1.3750 & 1.4038 & 1.3735 & 1.1172 \\
        Nova-Micro & 1.4098 & 1.3750 & 1.4054 & 1.3793 & 1.1713 \\
        Gemini 1.5 Pro & 1.5040 & 1.4722 & 1.5009 & 1.4609 & 1.1092 \\
        Gemini 1.5 Flash & 1.6594 & 1.6111 & 1.6516 & 1.6091 & 1.1839 \\
        Llama 3.3 (70B) & 1.7099 & 1.6528 & 1.6944 & 1.6497 & 1.2471 \\
        Mistral-Large-2407 & 1.7020 & 1.6528 & 1.7019 & 1.6491 & 1.2069 \\
        Grok 2 & 1.7366 & 1.6806 & 1.7301 & 1.6786 & 1.1621 \\
        Llama 3.1 (70B) & 1.7564 & 1.7083 & 1.7504 & 1.7003 & 1.1943 \\
        GPT-4 Turbo & 2.0990 & 2.0000 & 2.0792 & 1.9953 & 1.3299 \\
        \bottomrule
    \end{tabular}%
    }
\end{table}



\begin{table}[H]
    \caption{Relative cost of LLMs compared to canonical solutions on \textbf{Benchmark Set - II}.}
    \label{tab:relative-efficiency-b2}
    \centering
    \scriptsize
    \resizebox{\textwidth}{!}{%
    \begin{tabular}{lccccc}
        \toprule
        \textbf{Model} & \textbf{Avg. Package Energy} & \textbf{Avg. RAM Energy} & \textbf{Avg. Total Energy} & \textbf{Avg. Runtime} & \textbf{Avg. Memory} \\
        \midrule
        DeepSeek v3 & 1.1692 & 1.1493 & 1.1666 & 1.1599 & 1.0423 \\
        GPT-4o & 1.2171 & 1.1914 & 1.2147 & 1.2043 & 1.1236 \\
        Claude 3.5 Sonnet & 1.2361 & 1.2069 & 1.2347 & 1.2196 & 1.1069 \\
        Claude 3.5 Haiku & 1.2949 & 1.2644 & 1.2911 & 1.2768 & 1.1557 \\
        Gemini 2.0 Flash-Lite & 1.3428 & 1.3085 & 1.3388 & 1.3248 & 1.2496 \\
        Llama 3.1 (70B) & 1.6543 & 1.6026 & 1.6485 & 1.6145 & 1.2424 \\
        GPT-4 Turbo & 1.7293 & 1.6618 & 1.7211 & 1.6889 & 1.3722 \\
        Gemini 2.0 Flash & 1.8086 & 1.7206 & 1.7942 & 1.7491 & 1.1738 \\
        Llama 3.3 (70B) & 1.9356 & 1.8529 & 1.9258 & 1.8753 & 1.6210 \\
        Grok 2 & 2.0127 & 1.9260 & 2.0037 & 1.9516 & 1.4110 \\
        Gemini 1.5 Pro & 2.0545 & 1.9561 & 2.0431 & 1.9782 & 1.2755 \\
        \bottomrule
    \end{tabular}%
    }
\end{table}

\section{Problem Difficulty-Wise Code Energy Efficiency Analysis}
\label{Appendix:D}

The studied 878 Leetcode problems are divided into three categories (1) easy, (2) medium, and (3) hard problems based on the complexity of the algorithms or data structures required to solve the problems. In the dataset we use (detailed in Table\ref{table:benchmark_statistics}), there are 145 easy, 510 medium, and 223 hard problems. In this section, we present the detailed experimental results based on the problem difficulty levels. 

\subsection{ Benchmark Set - I: 20 LLMs \& 298 Common Problems}

\begin{table}[H]
    \caption{Performance and resource usage on \textbf{Easy} problems from \textbf{Benchmark Set - I}.}
    \label{tab:llm-benchmark-set-ii-prb-easy}
    \centering
    \scriptsize
    \resizebox{\textwidth}{!}{%
    \begin{tabular}{lccccc}
        \toprule
        \textbf{Model} & \textbf{Avg. Pkg Energy (J)} & \textbf{Avg. RAM Energy (J)} & \textbf{Avg. Total Energy (J)} & \textbf{Avg. Runtime (ms)} & \textbf{Avg. Mem (MB·s)} \\
        \midrule
        \textbf{Canonical Solution} & \cellcolor{gray!25}4.6447 & \cellcolor{gray!25}0.6672 & \cellcolor{gray!25}5.3119 & \cellcolor{gray!25}68.3146 & \cellcolor{gray!25}8.3332 \\
        Mistral-Large-2407 & 4.6200 & 0.6647 & 5.2847 & 68.3146 & 8.3324 \\
        Nova-Pro & 4.6311 & 0.6646 & 5.2957 & 67.8652 & 8.3681 \\
        Nova-Micro & 4.6408 & 0.6670 & 5.3078 & 68.0899 & 8.3968 \\
        Claude 3.5 Haiku & 4.6428 & 0.6674 & 5.3102 & 68.4270 & 8.3103 \\
        Pixtral-Large-2411 & 4.6430 & 0.6680 & 5.3110 & 68.0899 & 8.3804 \\
        DeepSeek v3 & 4.6451 & 0.6684 & 5.3135 & 68.3146 & 8.3545 \\
        Gemini 1.5 Pro & 4.6466 & 0.6681 & 5.3147 & 68.2022 & 8.3148 \\
        Codestral-Mamba-2407 & 4.6475 & 0.6674 & 5.3149 & 68.3146 & 8.3639 \\
        Llama 3.1 (70B) & 4.6471 & 0.6684 & 5.3155 & 68.3146 & 8.3431 \\
        Grok 2 & 4.6475 & 0.6688 & 5.3163 & 68.0899 & 8.3666 \\
        Nova-Lite & 4.6492 & 0.6687 & 5.3179 & 67.9775 & 8.3496 \\
        Gemini 2.0 Flash & 4.6525 & 0.6679 & 5.3203 & 68.0899 & 8.3389 \\
        GPT-4o & 4.6558 & 0.6666 & 5.3225 & 67.4157 & 8.3979 \\
        Gemini 2.0 Flash-Lite & 4.6570 & 0.6682 & 5.3252 & 68.2022 & 8.3786 \\
        Gemini 1.5 Flash & 4.6596 & 0.6700 & 5.3296 & 68.2022 & 8.4905 \\
        GPT-4 Turbo & 4.6616 & 0.6696 & 5.3311 & 68.2022 & 8.3600 \\
        GPT-3.5 Turbo & 4.7419 & 0.6800 & 5.4219 & 69.2135 & 8.4580 \\
        Claude 3.5 Sonnet & 5.3321 & 0.7569 & 6.0890 & 77.3034 & 8.7400 \\
        Llama 3.3 (70B) & 5.3324 & 0.7575 & 6.0899 & 77.5281 & 8.8151 \\
        Llama 3.1 (8B) & 5.4904 & 0.7793 & 6.2698 & 79.5506 & 8.8789 \\
        \bottomrule
    \end{tabular}%
    }
\end{table}

\begin{table}[H]
    \caption{Performance and resource usage on \textbf{Medium} problems from \textbf{Benchmark Set - I}.}
    \label{tab:llm-benchmark-set-i-prb-medium}
    \centering
    \scriptsize
    \resizebox{\textwidth}{!}{%
    \begin{tabular}{lccccc}
        \toprule
        \textbf{Model} & \textbf{Avg. Pkg Energy (J)} & \textbf{Avg. RAM Energy (J)} & \textbf{Avg. Total Energy (J)} & \textbf{Avg. Runtime (ms)} & \textbf{Avg. Mem (MB·s)} \\
        \midrule
        \textbf{Canonical Solution} & \cellcolor{gray!25}5.4982 & \cellcolor{gray!25}0.7878 & \cellcolor{gray!25}6.2860 & \cellcolor{gray!25}80.5028 & \cellcolor{gray!25}10.0253 \\
        Nova-Lite & 5.5962 & 0.7995 & 6.3957 & 81.4525 & 10.3356 \\
        DeepSeek v3 & 5.6428 & 0.8065 & 6.4493 & 82.5140 & 9.9460 \\
        Claude 3.5 Haiku & 5.8471 & 0.8351 & 6.6822 & 85.1397 & 10.6977 \\
        Nova-Pro & 5.8991 & 0.8411 & 6.7402 & 85.5866 & 10.8609 \\
        Gemini 2.0 Flash-Lite & 5.9157 & 0.8435 & 6.7592 & 86.0894 & 10.8333 \\
        GPT-3.5 Turbo & 5.9468 & 0.8491 & 6.7959 & 86.8156 & 10.9531 \\
        Gemini 2.0 Flash & 5.9648 & 0.8500 & 6.8148 & 86.5922 & 10.5324 \\
        Claude 3.5 Sonnet & 6.0437 & 0.8587 & 6.9023 & 87.3743 & 10.3755 \\
        Pixtral-Large-2411 & 6.3934 & 0.9042 & 7.2977 & 92.3464 & 10.1944 \\
        GPT-4o & 6.7210 & 0.9499 & 7.6709 & 96.8715 & 10.9429 \\
        Codestral-Mamba-2407 & 6.9753 & 0.9806 & 7.9559 & 100.7263 & 10.6409 \\
        Llama 3.1 (8B) & 7.0516 & 0.9901 & 8.0417 & 101.3966 & 10.9040 \\
        Nova-Micro & 7.3587 & 1.0334 & 8.3920 & 106.0894 & 11.9066 \\
        Llama 3.1 (70B) & 7.4142 & 1.0420 & 8.4562 & 106.3687 & 11.3072 \\
        Gemini 1.5 Pro & 8.1617 & 1.1371 & 9.2988 & 115.9777 & 11.0047 \\
        Gemini 1.5 Flash & 9.4430 & 1.3084 & 10.7514 & 134.3575 & 11.9996 \\
        Llama 3.3 (70B) & 9.5988 & 1.3296 & 10.9284 & 136.3128 & 12.8081 \\
        Mistral-Large-2407 & 9.8755 & 1.3639 & 11.2394 & 139.3296 & 12.4016 \\
        Grok 2 & 10.1078 & 1.3955 & 11.5033 & 143.0168 & 11.7598 \\
        GPT-4 Turbo & 13.0738 & 1.7823 & 14.8561 & 181.7877 & 14.1855 \\
        \bottomrule
    \end{tabular}%
    }
\end{table}

\begin{table}[H]
    \caption{Performance and resource usage on \textbf{Hard} problems from \textbf{Benchmark Set - I}.}
    \label{tab:llm-benchmark-set-ii-prb-hard}
    \centering
    \scriptsize
    \resizebox{\textwidth}{!}{%
    \begin{tabular}{lccccc}
        \toprule
        \textbf{Model} & \textbf{Avg. Pkg Energy (J)} & \textbf{Avg. RAM Energy (J)} & \textbf{Avg. Total Energy (J)} & \textbf{Avg. Runtime (ms)} & \textbf{Avg. Mem (MB·s)} \\
        \midrule
        \textbf{Canonical Solution} & \cellcolor{gray!25}3.6287 & \cellcolor{gray!25}0.5133 & \cellcolor{gray!25}4.1420 & \cellcolor{gray!25}53.6667 & \cellcolor{gray!25}1.8960 \\
        Gemini 2.0 Flash & 3.8263 & 0.5380 & 4.3643 & 56.6667 & 1.9679 \\
        Claude 3.5 Sonnet & 3.8613 & 0.5410 & 4.4023 & 56.0000 & 1.9753 \\
        DeepSeek v3 & 3.9007 & 0.5477 & 4.4483 & 56.3333 & 1.9838 \\
        GPT-4o & 4.9540 & 0.6860 & 5.6400 & 71.6667 & 2.3810 \\
        Llama 3.1 (8B) & 12.3187 & 1.6310 & 13.9497 & 171.0000 & 5.1774 \\
        Llama 3.3 (70B) & 12.3737 & 1.6427 & 14.0163 & 172.0000 & 5.2305 \\
        Grok 2 & 13.0240 & 1.7383 & 14.7623 & 181.3333 & 5.4729 \\
        Gemini 1.5 Flash & 13.0550 & 1.7417 & 14.7967 & 181.3333 & 5.4979 \\
        GPT-3.5 Turbo & 13.0660 & 1.7320 & 14.7980 & 182.3333 & 5.4916 \\
        GPT-4 Turbo & 13.0720 & 1.7373 & 14.8093 & 182.6667 & 5.5109 \\
        Gemini 1.5 Pro & 13.0773 & 1.7323 & 14.8097 & 181.6667 & 5.5012 \\
        Nova-Pro & 13.0793 & 1.7377 & 14.8170 & 182.3333 & 5.5022 \\
        Claude 3.5 Haiku & 13.0810 & 1.7400 & 14.8210 & 182.6667 & 5.5168 \\
        Nova-Lite & 13.0793 & 1.7423 & 14.8217 & 182.0000 & 5.5190 \\
        Nova-Micro & 13.0870 & 1.7380 & 14.8250 & 181.0000 & 5.5033 \\
        Mistral-Large-2407 & 13.0980 & 1.7320 & 14.8300 & 182.0000 & 5.5305 \\
        Codestral-Mamba-2407 & 13.1023 & 1.7413 & 14.8437 & 181.6667 & 5.4850 \\
        Gemini 2.0 Flash-Lite & 13.1087 & 1.7380 & 14.8467 & 182.3333 & 5.5093 \\
        Pixtral-Large-2411 & 13.1167 & 1.7473 & 14.8640 & 182.6667 & 5.5145 \\
        Llama 3.1 (70B) & 30.0567 & 3.9837 & 34.0403 & 415.0000 & 11.0132 \\
        \bottomrule
    \end{tabular}%
    }
\end{table}


\subsection{Benchmark Set - II: 11 LLMs \& 576 Common Problems}

\begin{table}[H]
    \caption{Performance and resource usage on \textbf{Easy} problems from \textbf{Benchmark Set - II}.}
    \label{tab:llm-benchmark-set-ii-prb-easy}
    \centering
    \scriptsize
    \resizebox{\textwidth}{!}{%
    \begin{tabular}{lccccc}
        \toprule
        \textbf{Model} & \textbf{Avg. Pkg Energy (J)} & \textbf{Avg. RAM Energy (J)} & \textbf{Avg. Total Energy (J)} & \textbf{Avg. Runtime (ms)} & \textbf{Avg. Mem (MB·s)} \\
        \midrule
        \textbf{Canonical Solution} & \cellcolor{gray!25}4.4834 & \cellcolor{gray!25}0.6386 & \cellcolor{gray!25}5.1220 & \cellcolor{gray!25}64.8739 & \cellcolor{gray!25}6.5104 \\
        Claude 3.5 Haiku & 4.4797 & 0.6379 & 5.1176 & 65.1261 & 6.5318 \\
        Grok 2 & 4.4999 & 0.6415 & 5.1414 & 65.2101 & 6.5511 \\
        DeepSeek v3 (37B) & 4.5446 & 0.6466 & 5.1913 & 65.7143 & 6.5245 \\
        Gemini 2.0 Flash & 4.5621 & 0.6474 & 5.2095 & 65.8824 & 6.5310 \\
        Gemini 1.5 Pro & 4.5654 & 0.6489 & 5.2143 & 65.7143 & 6.5549 \\
        GPT-4o & 4.5882 & 0.6517 & 5.2399 & 66.1345 & 6.5589 \\
        Gemini 2.0 Flash-Lite & 4.5886 & 0.6526 & 5.2412 & 66.4706 & 6.5649 \\
        GPT-4 Turbo & 4.5966 & 0.6534 & 5.2501 & 66.3025 & 6.5687 \\
        Llama 3.1 (70B) & 4.6179 & 0.6566 & 5.2745 & 66.3866 & 6.5949 \\
        Claude 3.5 Sonnet & 5.0890 & 0.7186 & 5.8076 & 73.0252 & 6.8626 \\
        Llama 3.3 (70B) & 5.1242 & 0.7218 & 5.8460 & 73.5294 & 6.8738 \\
        \bottomrule
    \end{tabular}%
    }
\end{table}

\begin{table}[H]
    \caption{Performance and resource usage on \textbf{Medium} problems from \textbf{Benchmark Set - II}.}
    \label{tab:llm-benchmark-set-ii-prb-medium}
    \centering
    \scriptsize
    \resizebox{\textwidth}{!}{%
    \begin{tabular}{lccccc}
        \toprule
        \textbf{Model} & \textbf{Avg. Pkg Energy (J)} & \textbf{Avg. RAM Energy (J)} & \textbf{Avg. Total Energy (J)} & \textbf{Avg. Runtime (ms)} & \textbf{Avg. Mem (MB·s)} \\
        \midrule
        \textbf{Canonical Solution} & \cellcolor{gray!25}5.1280 & \cellcolor{gray!25}0.7258 & \cellcolor{gray!25}5.8538 & \cellcolor{gray!25}74.1787 & \cellcolor{gray!25}7.7783 \\
        Claude 3.5 Haiku & 6.2490 & 0.8718 & 7.1208 & 89.5389 & 8.5283 \\
        DeepSeek v3 & 6.3454 & 0.8831 & 7.2285 & 90.8646 & 8.2535 \\
        Gemini 2.0 Flash-Lite & 6.4084 & 0.8924 & 7.3007 & 92.0749 & 8.8409 \\
        GPT-4o & 6.4459 & 0.8973 & 7.3432 & 92.3631 & 8.6318 \\
        Claude 3.5 Sonnet & 6.6971 & 0.9293 & 7.6264 & 95.5620 & 8.8311 \\
        Llama 3.1 (70B) & 7.7076 & 1.0615 & 8.7691 & 109.4236 & 9.1444 \\
        Llama 3.3 (70B) & 9.3395 & 1.2749 & 10.6145 & 131.6427 & 10.1128 \\
        GPT-4 Turbo & 9.5619 & 1.3073 & 10.8692 & 135.1297 & 10.3171 \\
        Gemini 2.0 Flash & 11.2547 & 1.5145 & 12.7692 & 156.9452 & 9.2581 \\
        Grok 2 & 11.5037 & 1.5532 & 13.0570 & 160.9510 & 9.7341 \\
        Gemini 1.5 Pro & 12.1494 & 1.6327 & 13.7821 & 168.9049 & 9.3706 \\
        \bottomrule
    \end{tabular}%
    }
\end{table}

\begin{table}[H]
    \caption{Performance and resource usage on \textbf{Hard} problems from \textbf{Benchmark Set - II}.}
    \label{tab:llm-benchmark-set-ii-prb-hard}
    \centering
    \scriptsize
    \resizebox{\textwidth}{!}{%
    \begin{tabular}{lccccc}
        \toprule
        \textbf{Model} & \textbf{Avg. Pkg Energy (J)} & \textbf{Avg. RAM Energy (J)} & \textbf{Avg. Total Energy (J)} & \textbf{Avg. Runtime (ms)} & \textbf{Avg. Mem (MB·s)} \\
        \midrule
        \textbf{Canonical Solution} & \cellcolor{gray!25}4.0241 & \cellcolor{gray!25}0.5670 & \cellcolor{gray!25}4.5911 & \cellcolor{gray!25}59.0000 & \cellcolor{gray!25}3.0738 \\
        Claude 3.5 Sonnet & 4.3294 & 0.6046 & 4.9340 & 62.5455 & 3.1147 \\
        DeepSeek v3 (37B) & 4.3489 & 0.6075 & 4.9565 & 63.5455 & 3.0378 \\
        Gemini 2.0 Flash & 4.7688 & 0.6631 & 5.4319 & 69.0000 & 4.3946 \\
        GPT-4o & 5.1903 & 0.7158 & 5.9061 & 74.5455 & 4.6173 \\
        Claude 3.5 Haiku & 7.8569 & 1.0631 & 8.9200 & 110.8182 & 6.0918 \\
        Gemini 1.5 Pro & 8.1489 & 1.0985 & 9.2474 & 114.1818 & 7.5059 \\
        GPT-4 Turbo & 8.1596 & 1.1035 & 9.2631 & 115.1818 & 7.8800 \\
        Gemini 2.0 Flash-Lite & 8.4450 & 1.1413 & 9.5863 & 118.9091 & 8.3057 \\
        Grok 2 & 9.2089 & 1.2430 & 10.4519 & 130.0909 & 11.1074 \\
        Llama 3.1 (70B) & 12.0945 & 1.6356 & 13.7301 & 169.5455 & 7.0872 \\
        Llama 3.3 (70B) & 13.4854 & 1.7825 & 15.2679 & 186.0000 & 16.8674 \\
        \bottomrule
    \end{tabular}%
    }
\end{table}

\section{Algorithm-Wise Code Energy Efficiency Analysis}
\label{Appendix:E}

The studied dataset includes twelve algorithmic methods, which are Greedy, Dynamic Programming (DP), Backtracking, Divide and Conquer, Depth-First Search (DFS), Breadth-First Search (BFS), Binary Search, Two Pointers, Sliding Window, Bit Manipulation, and Sorting (detailed in Table\ref{table:benchmark_statistics}). In this section, we present the detailed experimental results based on each algorithmic category. 

\subsection{ Benchmark Set - I: 20 LLMs \& 298 Common Problems}

\begin{table}[H]
    \caption{Performance and resource usage on \textbf{Greedy} problems from \textbf{Benchmark Set - I}.}
    \label{tab:llm-benchmark-set-i-prb-greedy}
    \centering
    \scriptsize
    \resizebox{\textwidth}{!}{%
    \begin{tabular}{lccccc}
        \toprule
        \textbf{Model} & \textbf{Avg. Pkg Energy (J)} & \textbf{Avg. RAM Energy (J)} & \textbf{Avg. Total Energy (J)} & \textbf{Avg. Runtime (ms)} & \textbf{Avg. Mem (MB·s)} \\
        \midrule
        \textbf{Canonical Solution} & \cellcolor{gray!25}4.1920 & \cellcolor{gray!25}0.6050 & \cellcolor{gray!25}4.7970 & \cellcolor{gray!25}61.5152 & \cellcolor{gray!25}7.5404 \\
        Pixtral-Large-2411 & 4.1347 & 0.5974 & 4.7321 & 60.7576 & 7.2395 \\
        Mistral-Large-2407 & 4.1361 & 0.5968 & 4.7329 & 60.4545 & 7.1168 \\
        Gemini 1.5 Pro & 4.1398 & 0.5979 & 4.7377 & 60.4545 & 7.2002 \\
        Grok 2 & 4.1426 & 0.5977 & 4.7403 & 60.4545 & 7.1998 \\
        DeepSeek v3 & 4.1420 & 0.5985 & 4.7405 & 60.7576 & 7.1993 \\
        Gemini 2.0 Flash & 4.1648 & 0.6009 & 4.7658 & 60.7576 & 7.3499 \\
        Claude 3.5 Haiku & 4.1885 & 0.6048 & 4.7933 & 61.3636 & 7.4778 \\
        GPT-4o & 4.1909 & 0.6036 & 4.7945 & 60.9091 & 7.4988 \\
        Codestral-Mamba-2407 & 4.2542 & 0.6112 & 4.8655 & 62.4242 & 7.9851 \\
        Claude 3.5 Sonnet & 4.2982 & 0.6185 & 4.9167 & 62.5758 & 8.0772 \\
        Gemini 2.0 Flash-Lite & 4.3129 & 0.6208 & 4.9336 & 63.0303 & 8.1648 \\
        Llama 3.1 (8B) & 4.3511 & 0.6271 & 4.9782 & 63.7879 & 8.4561 \\
        Llama 3.3 (70B) & 4.3515 & 0.6271 & 4.9786 & 63.7879 & 8.4575 \\
        Nova-Pro & 4.3588 & 0.6259 & 4.9847 & 63.3333 & 8.4165 \\
        Nova-Lite & 4.3591 & 0.6271 & 4.9862 & 63.1818 & 8.4439 \\
        Llama 3.1 (70B) & 4.3661 & 0.6289 & 4.9950 & 63.4848 & 8.4499 \\
        GPT-3.5 Turbo & 4.3691 & 0.6295 & 4.9986 & 63.9394 & 8.4216 \\
        Nova-Micro & 4.4288 & 0.6383 & 5.0671 & 65.3030 & 8.8569 \\
        Gemini 1.5 Flash & 4.4659 & 0.6411 & 5.1070 & 65.3030 & 8.4453 \\
        GPT-4 Turbo & 8.0168 & 1.1027 & 9.1195 & 113.9394 & 8.7471 \\
        \bottomrule
    \end{tabular}%
    }
\end{table}

\begin{table}[H]
    \caption{Performance and resource usage on \textbf{DP} problems from \textbf{Benchmark Set - I}.}
    \label{tab:llm-benchmark-set-i-prb-dp}
    \centering
    \scriptsize
    \resizebox{\textwidth}{!}{%
    \begin{tabular}{lccccc}
        \toprule
        \textbf{Model} & \textbf{Avg. Pkg Energy (J)} & \textbf{Avg. RAM Energy (J)} & \textbf{Avg. Total Energy (J)} & \textbf{Avg. Runtime (ms)} & \textbf{Avg. Mem (MB·s)} \\
        \midrule
        \textbf{Canonical Solution} & \cellcolor{gray!25}4.0514 & \cellcolor{gray!25}0.5678 & \cellcolor{gray!25}4.6192 & \cellcolor{gray!25}59.7297 & \cellcolor{gray!25}2.0289 \\
        DeepSeek v3 & 4.6659 & 0.6492 & 5.3151 & 68.2432 & 2.2492 \\
        Gemini 2.0 Flash & 5.2708 & 0.7312 & 6.0020 & 76.2162 & 3.5364 \\
        Claude 3.5 Sonnet & 5.3718 & 0.7416 & 6.1134 & 77.7027 & 2.4968 \\
        GPT-4o & 7.2693 & 0.9959 & 8.2653 & 103.3784 & 4.2553 \\
        Nova-Lite & 8.0426 & 1.0877 & 9.1303 & 113.9189 & 3.5552 \\
        Gemini 1.5 Pro & 8.0938 & 1.0900 & 9.1838 & 113.7838 & 3.5782 \\
        Pixtral-Large-2411 & 8.1832 & 1.1054 & 9.2886 & 115.4054 & 3.5840 \\
        GPT-4 Turbo & 8.2824 & 1.1161 & 9.3985 & 117.5676 & 3.6257 \\
        Gemini 2.0 Flash-Lite & 8.7188 & 1.1781 & 9.8969 & 123.5135 & 4.8863 \\
        Nova-Pro & 8.7930 & 1.1885 & 9.9815 & 124.4595 & 4.8438 \\
        Claude 3.5 Haiku & 8.8465 & 1.1959 & 10.0424 & 125.0000 & 4.9631 \\
        GPT-3.5 Turbo & 8.8945 & 1.2007 & 10.0951 & 125.8108 & 4.9758 \\
        Llama 3.3 (70B) & 10.1141 & 1.3597 & 11.4738 & 141.8919 & 5.3765 \\
        Mistral-Large-2407 & 10.6577 & 1.4330 & 12.0907 & 149.7297 & 5.5846 \\
        Gemini 1.5 Flash & 11.1853 & 1.5042 & 12.6895 & 157.0270 & 4.6200 \\
        Llama 3.1 (8B) & 11.2842 & 1.5046 & 12.7888 & 157.7027 & 4.6872 \\
        Codestral-Mamba-2407 & 11.4876 & 1.5392 & 13.0268 & 161.0811 & 4.6953 \\
        Nova-Micro & 11.7234 & 1.5686 & 13.2920 & 164.1892 & 4.7946 \\
        Llama 3.1 (70B) & 11.9942 & 1.6122 & 13.6064 & 168.3784 & 4.9080 \\
        Grok 2 & 13.1142 & 1.7599 & 14.8741 & 183.5135 & 5.2539 \\
        \bottomrule
    \end{tabular}%
    }
\end{table}

\begin{table}[H]
    \caption{Performance and resource usage on \textbf{Backtracking} problems from \textbf{Benchmark Set - I}.}
    \label{tab:llm-benchmark-set-i-prb-backtrack}
    \centering
    \scriptsize
    \resizebox{\textwidth}{!}{%
    \begin{tabular}{lccccc}
        \toprule
        \textbf{Model} & \textbf{Avg. Pkg Energy (J)} & \textbf{Avg. RAM Energy (J)} & \textbf{Avg. Total Energy (J)} & \textbf{Avg. Runtime (ms)} & \textbf{Avg. Mem (MB·s)} \\
        \midrule
        \textbf{Canonical Solution} & \cellcolor{gray!25}4.6818 & \cellcolor{gray!25}0.6509 & \cellcolor{gray!25}5.3327 & \cellcolor{gray!25}70.9091 & \cellcolor{gray!25}2.2505 \\
        Nova-Lite & 4.4455 & 0.6173 & 5.0627 & 63.6364 & 2.1453 \\
        Gemini 1.5 Pro & 4.5591 & 0.6318 & 5.1909 & 66.3636 & 2.2002 \\
        GPT-4 Turbo & 7.1027 & 0.9627 & 8.0655 & 101.8182 & 3.0507 \\
        DeepSeek v3 & 7.0991 & 0.9691 & 8.0682 & 101.8182 & 2.9887 \\
        Llama 3.1 (70B) & 7.4118 & 1.0091 & 8.4209 & 107.2727 & 3.1304 \\
        Nova-Pro & 9.3491 & 1.2864 & 10.6355 & 133.6364 & 10.8562 \\
        Gemini 2.0 Flash & 9.4955 & 1.3036 & 10.7991 & 135.4545 & 10.8023 \\
        Claude 3.5 Haiku & 9.7991 & 1.3509 & 11.1500 & 140.0000 & 11.6377 \\
        Llama 3.3 (70B) & 9.8800 & 1.3555 & 11.2355 & 140.0000 & 11.6134 \\
        Mistral-Large-2407 & 9.8855 & 1.3591 & 11.2445 & 140.0000 & 11.6881 \\
        GPT-3.5 Turbo & 9.9427 & 1.3655 & 11.3082 & 140.0000 & 11.5888 \\
        Gemini 2.0 Flash-Lite & 10.1800 & 1.3973 & 11.5773 & 144.5455 & 11.4508 \\
        Claude 3.5 Sonnet & 11.0155 & 1.4764 & 12.4918 & 153.6364 & 4.4692 \\
        GPT-4o & 12.6900 & 1.7182 & 14.4082 & 179.0909 & 12.4071 \\
        Pixtral-Large-2411 & 19.9027 & 2.6409 & 22.5436 & 277.2727 & 7.4959 \\
        Gemini 1.5 Flash & 25.2555 & 3.3664 & 28.6218 & 354.5455 & 9.1733 \\
        Codestral-Mamba-2407 & 27.3227 & 3.6236 & 30.9464 & 380.0000 & 9.8519 \\
        Nova-Micro & 27.9064 & 3.6991 & 31.6055 & 387.2727 & 10.1196 \\
        Llama 3.1 (8B) & 28.2391 & 3.7136 & 31.9527 & 389.0909 & 10.6508 \\
        Grok 2 & 28.9127 & 3.8382 & 32.7509 & 400.9091 & 10.4245 \\
        \bottomrule
    \end{tabular}%
    }
\end{table}

\begin{table}[H]
    \caption{Performance and resource usage on \textbf{Divide \& Conquer} problems from \textbf{Benchmark Set - I}.}
    \label{tab:llm-benchmark-set-i-prb-divideandconquer}
    \centering
    \scriptsize
    \resizebox{\textwidth}{!}{%
    \begin{tabular}{lccccc}
        \toprule
        \textbf{Model} & \textbf{Avg. Pkg Energy (J)} & \textbf{Avg. RAM Energy (J)} & \textbf{Avg. Total Energy (J)} & \textbf{Avg. Runtime (ms)} & \textbf{Avg. Mem (MB·s)} \\
        \midrule
        \textbf{Canonical Solution} & \cellcolor{gray!25}4.8525 & \cellcolor{gray!25}0.6750 & \cellcolor{gray!25}5.5275 & \cellcolor{gray!25}70.0000 & \cellcolor{gray!25}2.8257 \\
        Llama 3.1 (8B) & 4.6625 & 0.6500 & 5.3125 & 67.5000 & 2.7171 \\
        GPT-4o & 4.7700 & 0.6575 & 5.4275 & 70.0000 & 2.7633 \\
        Pixtral-Large-2411 & 4.9050 & 0.6750 & 5.5800 & 70.0000 & 2.8561 \\
        Nova-Micro & 5.0000 & 0.6975 & 5.6975 & 72.5000 & 2.8611 \\
        GPT-4 Turbo & 5.0250 & 0.6950 & 5.7200 & 77.5000 & 2.9864 \\
        Nova-Pro & 5.0600 & 0.7075 & 5.7675 & 75.0000 & 2.9898 \\
        Gemini 2.0 Flash & 5.1250 & 0.7125 & 5.8375 & 75.0000 & 2.9920 \\
        Claude 3.5 Sonnet & 5.1625 & 0.7100 & 5.8725 & 72.5000 & 2.9603 \\
        DeepSeek v3 & 5.1700 & 0.7150 & 5.8850 & 75.0000 & 3.0432 \\
        Gemini 1.5 Pro & 5.3100 & 0.7375 & 6.0475 & 77.5000 & 3.1163 \\
        Claude 3.5 Haiku & 5.3200 & 0.7425 & 6.0625 & 80.0000 & 3.1131 \\
        Llama 3.3 (70B) & 5.4025 & 0.7450 & 6.1475 & 77.5000 & 3.1728 \\
        Gemini 1.5 Flash & 5.4175 & 0.7475 & 6.1650 & 77.5000 & 3.1688 \\
        Nova-Lite & 5.4800 & 0.7525 & 6.2325 & 80.0000 & 3.2003 \\
        Codestral-Mamba-2407 & 5.4775 & 0.7550 & 6.2325 & 80.0000 & 3.1819 \\
        Mistral-Large-2407 & 5.6025 & 0.7675 & 6.3700 & 80.0000 & 3.2742 \\
        Grok 2 & 5.7350 & 0.7875 & 6.5225 & 82.5000 & 3.4081 \\
        Gemini 2.0 Flash-Lite & 5.7700 & 0.7925 & 6.5625 & 82.5000 & 3.3522 \\
        Llama 3.1 (70B) & 5.8300 & 0.8025 & 6.6325 & 85.0000 & 3.4416 \\
        GPT-3.5 Turbo & 6.8925 & 0.9375 & 7.8300 & 100.0000 & 4.0938 \\
        \bottomrule
    \end{tabular}%
    }
\end{table}

\begin{table}[H]
    \caption{Performance and resource usage on \textbf{DFS} problems from \textbf{Benchmark Set - I}.}
    \label{tab:llm-benchmark-set-i-prb-dfs}
    \centering
    \scriptsize
    \resizebox{\textwidth}{!}{%
    \begin{tabular}{lccccc}
        \toprule
        \textbf{Model} & \textbf{Avg. Pkg Energy (J)} & \textbf{Avg. RAM Energy (J)} & \textbf{Avg. Total Energy (J)} & \textbf{Avg. Runtime (ms)} & \textbf{Avg. Mem (MB·s)} \\
        \midrule
        \textbf{Canonical Solution} & \cellcolor{gray!25}4.4550 & \cellcolor{gray!25}0.6455 & \cellcolor{gray!25}5.1005 & \cellcolor{gray!25}69.0000 & \cellcolor{gray!25}3.5409 \\
        Gemini 2.0 Flash-Lite & 4.0690 & 0.5865 & 4.6555 & 62.0000 & 3.1474 \\
        GPT-4o & 4.0695 & 0.5875 & 4.6570 & 62.5000 & 3.1620 \\
        Pixtral-Large-2411 & 4.0715 & 0.5865 & 4.6580 & 62.5000 & 3.1565 \\
        GPT-4 Turbo & 4.0765 & 0.5875 & 4.6640 & 62.0000 & 3.1682 \\
        Llama 3.3 (70B) & 4.0760 & 0.5885 & 4.6645 & 62.5000 & 3.1440 \\
        Mistral-Large-2407 & 4.0775 & 0.5875 & 4.6650 & 62.5000 & 3.1600 \\
        Gemini 1.5 Flash & 4.0790 & 0.5885 & 4.6675 & 63.0000 & 3.1439 \\
        Codestral-Mamba-2407 & 4.0865 & 0.5885 & 4.6750 & 62.5000 & 3.1324 \\
        Llama 3.1 (70B) & 4.0890 & 0.5890 & 4.6780 & 62.5000 & 3.1718 \\
        GPT-3.5 Turbo & 4.0890 & 0.5890 & 4.6780 & 62.5000 & 3.1471 \\
        Gemini 2.0 Flash & 4.0915 & 0.5895 & 4.6810 & 62.0000 & 3.1579 \\
        Nova-Pro & 4.0965 & 0.5890 & 4.6855 & 62.0000 & 3.1583 \\
        Grok 2 & 4.0990 & 0.5890 & 4.6880 & 61.5000 & 3.1433 \\
        Nova-Lite & 4.1070 & 0.5920 & 4.6990 & 62.0000 & 3.1634 \\
        DeepSeek v3 & 4.1085 & 0.5910 & 4.6995 & 61.0000 & 3.1516 \\
        Gemini 1.5 Pro & 4.1085 & 0.5915 & 4.7000 & 61.5000 & 3.1616 \\
        Nova-Micro & 4.1135 & 0.5905 & 4.7040 & 61.5000 & 3.1423 \\
        Llama 3.1 (8B) & 4.1305 & 0.5930 & 4.7235 & 62.0000 & 3.1516 \\
        Claude 3.5 Haiku & 4.2700 & 0.6115 & 4.8815 & 64.5000 & 3.1951 \\
        Claude 3.5 Sonnet & 4.5315 & 0.6505 & 5.1820 & 67.0000 & 3.5600 \\
        \bottomrule
    \end{tabular}%
    }
\end{table}

\begin{table}[H]
    \caption{Performance and resource usage on \textbf{BFS} problems from \textbf{Benchmark Set - I}.}
    \label{tab:llm-benchmark-set-i-prb-bfs}
    \centering
    \scriptsize
    \resizebox{\textwidth}{!}{%
    \begin{tabular}{lccccc}
        \toprule
        \textbf{Model} & \textbf{Avg. Pkg Energy (J)} & \textbf{Avg. RAM Energy (J)} & \textbf{Avg. Total Energy (J)} & \textbf{Avg. Runtime (ms)} & \textbf{Avg. Mem (MB·s)} \\
        \midrule
        \textbf{Canonical Solution} & \cellcolor{gray!25}5.0531 & \cellcolor{gray!25}0.7281 & \cellcolor{gray!25}5.7812 & \cellcolor{gray!25}76.2500 & \cellcolor{gray!25}4.0769 \\
        Mistral-Large-2407 & 4.9769 & 0.7169 & 5.6938 & 75.0000 & 4.0199 \\
        Gemini 2.0 Flash-Lite & 4.9831 & 0.7169 & 5.7000 & 74.3750 & 4.0258 \\
        Llama 3.3 (70B) & 4.9819 & 0.7194 & 5.7012 & 75.0000 & 4.0185 \\
        GPT-3.5 Turbo & 4.9869 & 0.7169 & 5.7038 & 74.3750 & 3.9910 \\
        Gemini 1.5 Flash & 5.0075 & 0.7188 & 5.7263 & 74.3750 & 4.0083 \\
        Gemini 1.5 Pro & 5.0069 & 0.7200 & 5.7269 & 73.7500 & 4.0188 \\
        Gemini 2.0 Flash & 5.0094 & 0.7194 & 5.7288 & 74.3750 & 4.0175 \\
        Codestral-Mamba-2407 & 5.0131 & 0.7194 & 5.7325 & 75.0000 & 3.9936 \\
        Llama 3.1 (70B) & 5.0150 & 0.7206 & 5.7356 & 74.3750 & 4.0448 \\
        Nova-Lite & 5.0156 & 0.7213 & 5.7369 & 74.3750 & 4.0365 \\
        Nova-Micro & 5.0212 & 0.7188 & 5.7400 & 73.1250 & 4.0069 \\
        Llama 3.1 (8B) & 5.0288 & 0.7213 & 5.7500 & 74.3750 & 4.0200 \\
        Grok 2 & 5.0506 & 0.7231 & 5.7738 & 74.3750 & 4.0383 \\
        GPT-4 Turbo & 5.0575 & 0.7244 & 5.7819 & 74.3750 & 4.0674 \\
        Nova-Pro & 5.0712 & 0.7256 & 5.7969 & 75.0000 & 4.0737 \\
        GPT-4o & 5.0675 & 0.7325 & 5.8000 & 75.0000 & 4.0718 \\
        Claude 3.5 Haiku & 5.3544 & 0.7650 & 6.1194 & 78.7500 & 4.1307 \\
        DeepSeek v3 & 5.3781 & 0.7725 & 6.1506 & 78.7500 & 4.1291 \\
        Claude 3.5 Sonnet & 5.4075 & 0.7725 & 6.1800 & 80.0000 & 4.1781 \\
        Pixtral-Large-2411 & 5.7119 & 0.8150 & 6.5269 & 83.7500 & 4.2429 \\
        \bottomrule
    \end{tabular}%
    }
\end{table}

\begin{table}[H]
    \caption{Performance and resource usage on \textbf{Binary Search} problems from \textbf{Benchmark Set - I}.}
    \label{tab:llm-benchmark-set-i-prb-binarysearch}
    \centering
    \scriptsize
    \resizebox{\textwidth}{!}{%
    \begin{tabular}{lccccc}
        \toprule
        \textbf{Model} & \textbf{Avg. Pkg Energy (J)} & \textbf{Avg. RAM Energy (J)} & \textbf{Avg. Total Energy (J)} & \textbf{Avg. Runtime (ms)} & \textbf{Avg. Mem (MB·s)} \\
        \midrule
        \textbf{Canonical Solution} & \cellcolor{gray!25}10.9111 & \cellcolor{gray!25}1.5847 & \cellcolor{gray!25}12.4958 & \cellcolor{gray!25}158.4211 & \cellcolor{gray!25}35.3952 \\
        Mistral-Large-2407 & 10.8992 & 1.5853 & 12.4845 & 158.4211 & 35.4054 \\
        DeepSeek v3 & 10.9147 & 1.5858 & 12.5005 & 158.4211 & 35.4106 \\
        Gemini 2.0 Flash & 10.9171 & 1.5879 & 12.5050 & 159.2105 & 35.3539 \\
        Codestral-Mamba-2407 & 10.9192 & 1.5874 & 12.5066 & 158.4211 & 35.4220 \\
        Pixtral-Large-2411 & 10.9284 & 1.5892 & 12.5176 & 158.6842 & 35.4368 \\
        Claude 3.5 Sonnet & 10.9432 & 1.5863 & 12.5295 & 157.3684 & 35.2925 \\
        Nova-Lite & 10.9426 & 1.5871 & 12.5297 & 157.6316 & 35.5502 \\
        Nova-Pro & 10.9450 & 1.5866 & 12.5316 & 157.1053 & 35.4164 \\
        GPT-4o & 10.9468 & 1.5861 & 12.5329 & 156.5789 & 35.3753 \\
        Gemini 1.5 Pro & 10.9426 & 1.5911 & 12.5337 & 158.4211 & 35.4047 \\
        Claude 3.5 Haiku & 10.9529 & 1.5916 & 12.5445 & 158.6842 & 35.4130 \\
        Gemini 2.0 Flash-Lite & 10.9555 & 1.5897 & 12.5453 & 157.8947 & 35.3401 \\
        Grok 2 & 10.9550 & 1.5911 & 12.5461 & 157.8947 & 35.4858 \\
        Gemini 1.5 Flash & 10.9918 & 1.5987 & 12.5905 & 159.7368 & 35.6661 \\
        GPT-4 Turbo & 11.0218 & 1.5989 & 12.6208 & 159.2105 & 35.9505 \\
        GPT-3.5 Turbo & 11.1471 & 1.6184 & 12.7655 & 161.3158 & 35.7040 \\
        Llama 3.3 (70B) & 11.1613 & 1.6195 & 12.7808 & 161.8421 & 36.7364 \\
        Llama 3.1 (8B) & 11.2816 & 1.6371 & 12.9187 & 163.4211 & 35.8710 \\
        Nova-Micro & 11.5818 & 1.6789 & 13.2608 & 167.3684 & 39.5061 \\
        Llama 3.1 (70B) & 24.5637 & 3.3911 & 27.9547 & 345.5263 & 40.9890 \\
        \bottomrule
    \end{tabular}%
    }
\end{table}

\begin{table}[H]
    \caption{Performance and resource usage on \textbf{Two Pointers} problems from \textbf{Benchmark Set - I}.}
    \label{tab:llm-benchmark-set-i-prb-twopointers}
    \centering
    \scriptsize
    \resizebox{\textwidth}{!}{%
    \begin{tabular}{lccccc}
        \toprule
        \textbf{Model} & \textbf{Avg. Pkg Energy (J)} & \textbf{Avg. RAM Energy (J)} & \textbf{Avg. Total Energy (J)} & \textbf{Avg. Runtime (ms)} & \textbf{Avg. Mem (MB·s)} \\
        \midrule
        \textbf{Canonical Solution} & \cellcolor{gray!25}6.7030 & \cellcolor{gray!25}0.9758 & \cellcolor{gray!25}7.6788 & \cellcolor{gray!25}97.8000 & \cellcolor{gray!25}21.1321 \\
        Grok 2 & 6.6762 & 0.9720 & 7.6482 & 96.8000 & 20.7135 \\
        DeepSeek v3 & 6.6846 & 0.9726 & 7.6572 & 97.4000 & 20.6268 \\
        Mistral-Large-2407 & 6.7170 & 0.9768 & 7.6938 & 97.6000 & 20.6131 \\
        Gemini 1.5 Pro & 6.7312 & 0.9786 & 7.7098 & 97.4000 & 20.7130 \\
        Claude 3.5 Haiku & 6.7298 & 0.9806 & 7.7104 & 98.2000 & 21.0926 \\
        Gemini 2.0 Flash & 6.7488 & 0.9818 & 7.7306 & 98.2000 & 20.8350 \\
        GPT-4o & 6.8024 & 0.9872 & 7.7896 & 98.0000 & 20.9696 \\
        Pixtral-Large-2411 & 6.8050 & 0.9884 & 7.7934 & 98.4000 & 20.7085 \\
        Codestral-Mamba-2407 & 6.8074 & 0.9866 & 7.7940 & 98.4000 & 21.6941 \\
        Claude 3.5 Sonnet & 6.8930 & 0.9990 & 7.8920 & 99.6000 & 21.7388 \\
        GPT-3.5 Turbo & 6.9266 & 1.0064 & 7.9330 & 100.6000 & 22.2774 \\
        Nova-Pro & 6.9366 & 1.0064 & 7.9430 & 100.0000 & 22.2237 \\
        Nova-Lite & 6.9404 & 1.0052 & 7.9456 & 100.4000 & 22.3925 \\
        Llama 3.1 (8B) & 6.9444 & 1.0092 & 7.9536 & 100.6000 & 22.4614 \\
        Gemini 2.0 Flash-Lite & 6.9490 & 1.0072 & 7.9562 & 100.4000 & 21.9128 \\
        Gemini 1.5 Flash & 6.9918 & 1.0136 & 8.0054 & 101.8000 & 22.1877 \\
        GPT-4 Turbo & 7.0012 & 1.0136 & 8.0148 & 101.4000 & 21.5005 \\
        Llama 3.3 (70B) & 7.2122 & 1.0442 & 8.2564 & 104.4000 & 23.2472 \\
        Llama 3.1 (70B) & 7.2118 & 1.0466 & 8.2584 & 104.2000 & 23.3472 \\
        Nova-Micro & 7.6618 & 1.1070 & 8.7688 & 110.6000 & 25.9528 \\
        \bottomrule
    \end{tabular}%
    }
\end{table}

\begin{table}[H]
    \caption{Performance and resource usage on \textbf{Sliding Window} problems from \textbf{Benchmark Set - I}.}
    \label{tab:llm-benchmark-set-i-prb-slidingwindow}
    \centering
    \scriptsize
    \resizebox{\textwidth}{!}{%
    \begin{tabular}{lccccc}
        \toprule
        \textbf{Model} & \textbf{Avg. Pkg Energy (J)} & \textbf{Avg. RAM Energy (J)} & \textbf{Avg. Total Energy (J)} & \textbf{Avg. Runtime (ms)} & \textbf{Avg. Mem (MB·s)} \\
        \midrule
        \textbf{Canonical Solution} & \cellcolor{gray!25}7.3565 & \cellcolor{gray!25}1.0691 & \cellcolor{gray!25}8.4257 & \cellcolor{gray!25}105.2174 & \cellcolor{gray!25}26.9084 \\
        GPT-3.5 Turbo & 7.3248 & 1.0700 & 8.3948 & 106.5217 & 26.8932 \\
        Mistral-Large-2407 & 7.3365 & 1.0704 & 8.4070 & 106.0870 & 26.9404 \\
        DeepSeek v3 & 7.3435 & 1.0687 & 8.4122 & 106.0870 & 26.7775 \\
        Pixtral-Large-2411 & 7.3435 & 1.0717 & 8.4152 & 106.0870 & 26.7335 \\
        GPT-4o & 7.3483 & 1.0683 & 8.4165 & 105.2174 & 26.6089 \\
        Gemini 2.0 Flash & 7.3465 & 1.0700 & 8.4165 & 106.0870 & 26.7481 \\
        Nova-Pro & 7.3483 & 1.0717 & 8.4200 & 105.6522 & 26.7771 \\
        Claude 3.5 Haiku & 7.3530 & 1.0726 & 8.4257 & 105.6522 & 26.9961 \\
        Claude 3.5 Sonnet & 7.3565 & 1.0696 & 8.4261 & 106.0870 & 26.6658 \\
        Llama 3.1 (8B) & 7.3617 & 1.0739 & 8.4357 & 105.6522 & 27.1564 \\
        Gemini 2.0 Flash-Lite & 7.3709 & 1.0717 & 8.4426 & 106.0870 & 26.7718 \\
        Codestral-Mamba-2407 & 7.3752 & 1.0730 & 8.4483 & 105.2174 & 26.8745 \\
        Gemini 1.5 Pro & 7.3809 & 1.0739 & 8.4548 & 106.0870 & 26.9190 \\
        Gemini 1.5 Flash & 7.3839 & 1.0739 & 8.4578 & 105.2174 & 26.6743 \\
        Nova-Lite & 7.3870 & 1.0722 & 8.4591 & 106.0870 & 27.0621 \\
        Grok 2 & 7.3961 & 1.0752 & 8.4713 & 105.6522 & 26.9730 \\
        GPT-4 Turbo & 7.4917 & 1.0904 & 8.5822 & 108.2609 & 27.7598 \\
        Llama 3.3 (70B) & 7.7465 & 1.1243 & 8.8709 & 111.3043 & 28.7879 \\
        Llama 3.1 (70B) & 7.7552 & 1.1283 & 8.8835 & 110.8696 & 29.0314 \\
        Nova-Micro & 8.4552 & 1.2248 & 9.6800 & 121.3043 & 33.5166 \\
        \bottomrule
    \end{tabular}%
    }
\end{table}

\begin{table}[H]
    \caption{Performance and resource usage on \textbf{Bit Manipulation} problems from \textbf{Benchmark Set - I}.}
    \label{tab:llm-benchmark-set-i-prb-bitmanipulation}
    \centering
    \scriptsize
    \resizebox{\textwidth}{!}{%
    \begin{tabular}{lccccc}
        \toprule
        \textbf{Model} & \textbf{Avg. Pkg Energy (J)} & \textbf{Avg. RAM Energy (J)} & \textbf{Avg. Total Energy (J)} & \textbf{Avg. Runtime (ms)} & \textbf{Avg. Mem (MB·s)} \\
        \midrule
        \textbf{Canonical Solution} & \cellcolor{gray!25}3.0623 & \cellcolor{gray!25}0.4354 & \cellcolor{gray!25}3.4977 & \cellcolor{gray!25}45.1429 & \cellcolor{gray!25}1.6504 \\
        Claude 3.5 Haiku & 2.9423 & 0.4203 & 3.3626 & 44.0000 & 1.6011 \\
        GPT-4o & 2.9443 & 0.4209 & 3.3651 & 44.2857 & 1.6155 \\
        DeepSeek v3 & 2.9451 & 0.4211 & 3.3663 & 44.2857 & 1.6093 \\
        Mistral-Large-2407 & 2.9503 & 0.4200 & 3.3703 & 43.7143 & 1.6112 \\
        Claude 3.5 Sonnet & 2.9514 & 0.4209 & 3.3723 & 43.7143 & 1.6081 \\
        Codestral-Mamba-2407 & 2.9826 & 0.4249 & 3.4074 & 44.2857 & 1.6116 \\
        GPT-3.5 Turbo & 2.9871 & 0.4257 & 3.4129 & 44.0000 & 1.6220 \\
        Gemini 2.0 Flash & 2.9974 & 0.4271 & 3.4246 & 44.2857 & 1.6277 \\
        Gemini 2.0 Flash-Lite & 3.0003 & 0.4257 & 3.4260 & 44.0000 & 1.6259 \\
        Nova-Micro & 3.0114 & 0.4294 & 3.4409 & 44.5714 & 1.6232 \\
        GPT-4 Turbo & 3.0380 & 0.4337 & 3.4717 & 45.1429 & 1.6416 \\
        Nova-Pro & 3.0654 & 0.4360 & 3.5014 & 45.4286 & 1.6584 \\
        Nova-Lite & 3.0689 & 0.4360 & 3.5049 & 44.8571 & 1.6566 \\
        Llama 3.1 (8B) & 3.1029 & 0.4429 & 3.5457 & 46.2857 & 1.6706 \\
        Llama 3.1 (70B) & 3.1137 & 0.4417 & 3.5554 & 45.4286 & 1.6677 \\
        Pixtral-Large-2411 & 7.1914 & 0.9760 & 8.1674 & 101.7143 & 3.0908 \\
        Grok 2 & 15.6151 & 2.0931 & 17.7083 & 216.2857 & 7.5604 \\
        Gemini 1.5 Flash & 15.6614 & 2.1029 & 17.7643 & 216.8571 & 7.6045 \\
        Gemini 1.5 Pro & 16.4140 & 2.1906 & 18.6046 & 227.1429 & 7.3323 \\
        Llama 3.3 (70B) & 18.2411 & 2.4380 & 20.6791 & 252.2857 & 8.6808 \\
        \bottomrule
    \end{tabular}%
    }
\end{table}

\begin{table}[H]
    \caption{Performance and resource usage on \textbf{Sorting} problems from \textbf{Benchmark Set - I}.}
    \label{tab:llm-benchmark-set-i-prb-sorting}
    \centering
    \scriptsize
    \resizebox{\textwidth}{!}{%
    \begin{tabular}{lccccc}
        \toprule
        \textbf{Model} & \textbf{Avg. Pkg Energy (J)} & \textbf{Avg. RAM Energy (J)} & \textbf{Avg. Total Energy (J)} & \textbf{Avg. Runtime (ms)} & \textbf{Avg. Mem (MB·s)} \\
        \midrule
        \textbf{Canonical Solution} & \cellcolor{gray!25}6.5265 & \cellcolor{gray!25}0.9441 & \cellcolor{gray!25}7.4706 & \cellcolor{gray!25}94.8235 & \cellcolor{gray!25}17.3111 \\
        Pixtral-Large-2411 & 6.5282 & 0.9449 & 7.4732 & 94.8235 & 17.0698 \\
        DeepSeek v3 & 6.5451 & 0.9465 & 7.4915 & 95.4118 & 17.0882 \\
        Gemini 2.0 Flash & 6.6494 & 0.9607 & 7.6101 & 96.4706 & 17.1853 \\
        Gemini 1.5 Pro & 6.6761 & 0.9633 & 7.6394 & 96.4706 & 17.0801 \\
        GPT-4o & 6.6835 & 0.9633 & 7.6468 & 96.0000 & 17.4976 \\
        Grok 2 & 6.6864 & 0.9659 & 7.6522 & 97.0588 & 17.1309 \\
        Claude 3.5 Haiku & 6.6978 & 0.9691 & 7.6668 & 97.6471 & 17.6077 \\
        Codestral-Mamba-2407 & 6.7938 & 0.9780 & 7.7718 & 98.8235 & 17.7396 \\
        Nova-Pro & 6.8233 & 0.9831 & 7.8064 & 98.5882 & 18.0333 \\
        Gemini 2.0 Flash-Lite & 6.8932 & 0.9925 & 7.8856 & 99.8824 & 17.9685 \\
        GPT-3.5 Turbo & 6.9145 & 0.9968 & 7.9113 & 100.4706 & 18.1414 \\
        Nova-Lite & 6.9166 & 0.9960 & 7.9126 & 100.1176 & 18.0759 \\
        Gemini 1.5 Flash & 7.0067 & 1.0073 & 8.0140 & 101.4118 & 18.1709 \\
        Nova-Micro & 7.2829 & 1.0486 & 8.3315 & 105.8824 & 20.2935 \\
        Claude 3.5 Sonnet & 7.4954 & 1.0695 & 8.5649 & 107.5294 & 18.1460 \\
        Llama 3.1 (8B) & 7.5456 & 1.0794 & 8.6251 & 108.8235 & 18.5860 \\
        Llama 3.3 (70B) & 7.6629 & 1.0959 & 8.7588 & 110.5882 & 19.0995 \\
        Llama 3.1 (70B) & 9.5380 & 1.3485 & 10.8865 & 136.2353 & 19.7630 \\
        Mistral-Large-2407 & 13.6091 & 1.8732 & 15.4822 & 190.1176 & 20.6693 \\
        GPT-4 Turbo & 22.2915 & 3.0160 & 25.3075 & 306.0000 & 26.0765 \\
        \bottomrule
    \end{tabular}%
    }
\end{table}

\subsection{Benchmark Set - II: 11 LLMs \& 576 Common Problems}

\begin{table}[H]
    \caption{Performance and resource usage on \textbf{Greedy} problems from \textbf{Benchmark Set - II}.}
    \label{tab:llm-benchmark-set-ii-prb-greedy}
    \centering
    \scriptsize
    \resizebox{\textwidth}{!}{%
    \begin{tabular}{lccccc}
        \toprule
        \textbf{Model} & \textbf{Avg. Pkg Energy (J)} & \textbf{Avg. RAM Energy (J)} & \textbf{Avg. Total Energy (J)} & \textbf{Avg. Runtime (ms)} & \textbf{Avg. Mem (MB·s)} \\
        \midrule
        \textbf{Canonical Solution} & \cellcolor{gray!25}3.6603 & \cellcolor{gray!25}0.5175 & \cellcolor{gray!25}4.1778 & \cellcolor{gray!25}53.2215 & \cellcolor{gray!25}4.1165 \\
        Grok 2 & 3.6160 & 0.5127 & 4.1287 & 52.8859 & 3.9836 \\
        DeepSeek v3 (37B) & 3.6323 & 0.5140 & 4.1464 & 52.4832 & 3.9999 \\
        GPT-4o & 3.6930 & 0.5221 & 4.2150 & 53.6242 & 4.1181 \\
        Gemini 1.5 Pro & 3.7479 & 0.5287 & 4.2765 & 54.2282 & 4.0302 \\
        Claude 3.5 Sonnet & 3.8555 & 0.5423 & 4.3978 & 55.5034 & 4.4288 \\
        Claude 3.5 Haiku & 3.8604 & 0.5430 & 4.4034 & 55.4362 & 4.1778 \\
        Llama 3.1 (70B) & 3.8611 & 0.5439 & 4.4050 & 56.0403 & 4.6114 \\
        Gemini 2.0 Flash-Lite & 3.8744 & 0.5455 & 4.4199 & 56.0403 & 4.4935 \\
        Gemini 2.0 Flash & 3.8809 & 0.5462 & 4.4272 & 56.1745 & 4.1288 \\
        Llama 3.3 (70B) & 3.9405 & 0.5539 & 4.4944 & 57.1141 & 4.6310 \\
        GPT-4 Turbo & 5.3377 & 0.7378 & 6.0755 & 76.4430 & 4.6970 \\
        \bottomrule
    \end{tabular}%
    }
\end{table}

\begin{table}[H]
    \caption{Performance and resource usage on \textbf{DP} problems from \textbf{Benchmark Set - II}.}
    \label{tab:llm-benchmark-set-ii-prb-dynamicprogramming}
    \centering
    \scriptsize
    \resizebox{\textwidth}{!}{%
    \begin{tabular}{lccccc}
        \toprule
        \textbf{Model} & \textbf{Avg. Pkg Energy (J)} & \textbf{Avg. RAM Energy (J)} & \textbf{Avg. Total Energy (J)} & \textbf{Avg. Runtime (ms)} & \textbf{Avg. Mem (MB·s)} \\
        \midrule
        \textbf{Canonical Solution} & \cellcolor{gray!25}4.1146 & \cellcolor{gray!25}0.5721 & \cellcolor{gray!25}4.6867 & \cellcolor{gray!25}59.6154 & \cellcolor{gray!25}2.7721 \\
        DeepSeek v3 (37B) & 6.3665 & 0.8607 & 7.2272 & 90.3205 & 3.8131 \\
        GPT-4o & 6.9045 & 0.9315 & 7.8360 & 97.2436 & 5.4728 \\
        Claude 3.5 Sonnet & 7.0043 & 0.9426 & 7.9469 & 98.6538 & 4.7442 \\
        GPT-4 Turbo & 7.8680 & 1.0578 & 8.9258 & 110.2564 & 6.7915 \\
        Claude 3.5 Haiku & 8.3847 & 1.1260 & 9.5108 & 117.6923 & 6.3315 \\
        Gemini 2.0 Flash-Lite & 9.2200 & 1.2338 & 10.4538 & 129.0385 & 8.3407 \\
        Llama 3.1 (70B) & 11.0878 & 1.4747 & 12.5625 & 153.8462 & 6.4045 \\
        Llama 3.3 (70B) & 15.7245 & 2.0661 & 17.7906 & 216.1538 & 15.1335 \\
        Gemini 2.0 Flash & 17.4983 & 2.2915 & 19.7897 & 239.6795 & 6.8631 \\
        Grok 2 & 18.2125 & 2.3948 & 20.6073 & 251.3462 & 11.3735 \\
        Gemini 1.5 Pro & 18.7244 & 2.4491 & 21.1735 & 256.2821 & 8.0891 \\
        \bottomrule
    \end{tabular}%
    }
\end{table}

\begin{table}[H]
    \caption{Performance and resource usage on \textbf{Backtracking} problems from \textbf{Benchmark Set - II}.}
    \label{tab:llm-benchmark-set-ii-prb-backtrack}
    \centering
    \scriptsize
    \resizebox{\textwidth}{!}{%
    \begin{tabular}{lccccc}
        \toprule
        \textbf{Model} & \textbf{Avg. Pkg Energy (J)} & \textbf{Avg. RAM Energy (J)} & \textbf{Avg. Total Energy (J)} & \textbf{Avg. Runtime (ms)} & \textbf{Avg. Mem (MB·s)} \\
        \midrule
        \textbf{Canonical Solution} & \cellcolor{gray!25}8.9544 & \cellcolor{gray!25}1.2652 & \cellcolor{gray!25}10.2196 & \cellcolor{gray!25}127.2000 & \cellcolor{gray!25}21.9081 \\
        GPT-4 Turbo & 11.8400 & 1.6328 & 13.4728 & 166.4000 & 23.5252 \\
        GPT-4o & 15.0500 & 2.0664 & 17.1164 & 210.8000 & 27.6705 \\
        Claude 3.5 Sonnet & 15.3848 & 2.0932 & 17.4780 & 216.0000 & 24.7516 \\
        Gemini 2.0 Flash-Lite & 17.1360 & 2.3276 & 19.4636 & 239.2000 & 28.3890 \\
        Claude 3.5 Haiku & 17.5416 & 2.3788 & 19.9204 & 244.0000 & 28.2667 \\
        DeepSeek v3 (37B) & 19.0252 & 2.5568 & 21.5820 & 264.0000 & 25.6659 \\
        Llama 3.1 (70B) & 24.5892 & 3.2872 & 27.8764 & 340.4000 & 27.7724 \\
        Llama 3.3 (70B) & 55.8324 & 7.2816 & 63.1140 & 756.8000 & 74.2700 \\
        Grok 2 & 66.9000 & 8.7356 & 75.6356 & 913.6000 & 36.2330 \\
        Gemini 1.5 Pro & 79.7148 & 10.3496 & 90.0644 & 1077.6000 & 32.5344 \\
        Gemini 2.0 Flash & 82.4440 & 10.7056 & 93.1496 & 1114.4000 & 36.8112 \\
        \bottomrule
    \end{tabular}%
    }
\end{table}

\begin{table}[H]
    \caption{Performance and resource usage on \textbf{Divide \& Conquer} problems from \textbf{Benchmark Set - II}.}
    \label{tab:llm-benchmark-set-ii-prb-divideandconquer}
    \centering
    \scriptsize
    \resizebox{\textwidth}{!}{%
    \begin{tabular}{lccccc}
        \toprule
        \textbf{Model} & \textbf{Avg. Pkg Energy (J)} & \textbf{Avg. RAM Energy (J)} & \textbf{Avg. Total Energy (J)} & \textbf{Avg. Runtime (ms)} & \textbf{Avg. Mem (MB·s)} \\
        \midrule
        \textbf{Canonical Solution} & \cellcolor{gray!25}3.8542 & \cellcolor{gray!25}0.5350 & \cellcolor{gray!25}4.3892 & \cellcolor{gray!25}55.8333 & \cellcolor{gray!25}2.0303 \\
        Claude 3.5 Haiku & 3.8808 & 0.5375 & 4.4183 & 56.6667 & 2.0862 \\
        Grok 2 & 3.9375 & 0.5450 & 4.4825 & 56.6667 & 2.1585 \\
        DeepSeek v3 (37B) & 4.4425 & 0.6108 & 5.0533 & 64.1667 & 2.2794 \\
        Claude 3.5 Sonnet & 4.4492 & 0.6117 & 5.0608 & 64.1667 & 2.2634 \\
        GPT-4 Turbo & 4.4583 & 0.6142 & 5.0725 & 65.0000 & 2.2651 \\
        Gemini 2.0 Flash & 4.5067 & 0.6192 & 5.1258 & 65.8333 & 2.2838 \\
        Llama 3.3 (70B) & 4.5083 & 0.6217 & 5.1300 & 65.0000 & 2.3204 \\
        Gemini 1.5 Pro & 4.5183 & 0.6192 & 5.1375 & 64.1667 & 2.2972 \\
        GPT-4o & 4.5700 & 0.6267 & 5.1967 & 66.6667 & 2.2528 \\
        Gemini 2.0 Flash-Lite & 4.7267 & 0.6458 & 5.3725 & 66.6667 & 2.4001 \\
        Llama 3.1 (70B) & 4.7467 & 0.6517 & 5.3983 & 69.1667 & 2.4406 \\
        \bottomrule
    \end{tabular}%
    }
\end{table}

\begin{table}[H]
    \caption{Performance and resource usage on \textbf{DFS} problems from \textbf{Benchmark Set - II}.}
    \label{tab:llm-benchmark-set-ii-prb-dfs}
    \centering
    \scriptsize
    \resizebox{\textwidth}{!}{%
    \begin{tabular}{lccccc}
        \toprule
        \textbf{Model} & \textbf{Avg. Pkg Energy (J)} & \textbf{Avg. RAM Energy (J)} & \textbf{Avg. Total Energy (J)} & \textbf{Avg. Runtime (ms)} & \textbf{Avg. Mem (MB·s)} \\
        \midrule
        \textbf{Canonical Solution} & \cellcolor{gray!25}6.1661 & \cellcolor{gray!25}0.8848 & \cellcolor{gray!25}7.0509 & \cellcolor{gray!25}91.5152 & \cellcolor{gray!25}6.8849 \\
        Grok 2 & 5.9167 & 0.8482 & 6.7648 & 88.1818 & 6.6496 \\
        Gemini 1.5 Pro & 5.9215 & 0.8482 & 6.7697 & 87.8788 & 6.6613 \\
        DeepSeek v3 (37B) & 5.9264 & 0.8470 & 6.7733 & 87.5758 & 6.6432 \\
        GPT-4 Turbo & 5.9336 & 0.8503 & 6.7839 & 88.1818 & 6.6526 \\
        GPT-4o & 5.9548 & 0.8509 & 6.8058 & 88.4848 & 6.6690 \\
        Gemini 2.0 Flash-Lite & 5.9630 & 0.8542 & 6.8173 & 89.3939 & 6.6983 \\
        Llama 3.1 (70B) & 5.9776 & 0.8545 & 6.8321 & 88.7879 & 6.6550 \\
        Gemini 2.0 Flash & 5.9885 & 0.8564 & 6.8448 & 88.1818 & 6.7327 \\
        Claude 3.5 Haiku & 6.0476 & 0.8639 & 6.9115 & 90.3030 & 6.7167 \\
        Llama 3.3 (70B) & 6.0609 & 0.8667 & 6.9276 & 89.3939 & 6.7028 \\
        Claude 3.5 Sonnet & 6.1588 & 0.8827 & 7.0415 & 90.9091 & 6.8542 \\
        \bottomrule
    \end{tabular}%
    }
\end{table}

\begin{table}[H]
    \caption{Performance and resource usage on \textbf{BFS} problems from \textbf{Benchmark Set - II}.}
    \label{tab:llm-benchmark-set-ii-prb-bfs}
    \centering
    \scriptsize
    \resizebox{\textwidth}{!}{%
    \begin{tabular}{lccccc}
        \toprule
        \textbf{Model} & \textbf{Avg. Pkg Energy (J)} & \textbf{Avg. RAM Energy (J)} & \textbf{Avg. Total Energy (J)} & \textbf{Avg. Runtime (ms)} & \textbf{Avg. Mem (MB·s)} \\
        \midrule
        \textbf{Canonical Solution} & \cellcolor{gray!25}7.6440 & \cellcolor{gray!25}1.0747 & \cellcolor{gray!25}8.7187 & \cellcolor{gray!25}110.6667 & \cellcolor{gray!25}7.9833 \\
        Gemini 2.0 Flash & 10.1163 & 1.4137 & 11.5300 & 148.0000 & 7.9332 \\
        Gemini 2.0 Flash-Lite & 10.1893 & 1.4207 & 11.6100 & 148.3333 & 7.9233 \\
        Llama 3.3 (70B) & 10.2653 & 1.4310 & 11.6963 & 149.0000 & 7.9710 \\
        GPT-4o & 10.5053 & 1.4613 & 11.9667 & 152.0000 & 8.0799 \\
        Claude 3.5 Sonnet & 10.6250 & 1.4807 & 12.1057 & 154.0000 & 8.0730 \\
        GPT-4 Turbo & 10.6593 & 1.4837 & 12.1430 & 154.6667 & 8.1395 \\
        Gemini 1.5 Pro & 10.6823 & 1.4873 & 12.1697 & 154.0000 & 8.1141 \\
        Grok 2 & 10.7960 & 1.5027 & 12.2987 & 156.0000 & 8.1800 \\
        DeepSeek v3 (37B) & 10.8820 & 1.5113 & 12.3933 & 157.0000 & 8.1797 \\
        Claude 3.5 Haiku & 10.9117 & 1.5113 & 12.4230 & 157.6667 & 8.2380 \\
        Llama 3.1 (70B) & 11.0823 & 1.5393 & 12.6217 & 160.0000 & 8.3741 \\
        \bottomrule
    \end{tabular}%
    }
\end{table}

\begin{table}[H]
    \caption{Performance and resource usage on \textbf{Binary Search} problems from \textbf{Benchmark Set - II}.}
    \label{tab:llm-benchmark-set-ii-prb-binarysearch}
    \centering
    \scriptsize
    \resizebox{\textwidth}{!}{%
    \begin{tabular}{lccccc}
        \toprule
        \textbf{Model} & \textbf{Avg. Pkg Energy (J)} & \textbf{Avg. RAM Energy (J)} & \textbf{Avg. Total Energy (J)} & \textbf{Avg. Runtime (ms)} & \textbf{Avg. Mem (MB·s)} \\
        \midrule
        \textbf{Canonical Solution} & \cellcolor{gray!25}7.4294 & \cellcolor{gray!25}1.0772 & \cellcolor{gray!25}8.5066 & \cellcolor{gray!25}108.3721 & \cellcolor{gray!25}17.6033 \\
        DeepSeek v3 (37B) & 7.3084 & 1.0566 & 8.3650 & 106.2791 & 17.2535 \\
        Claude 3.5 Sonnet & 7.3731 & 1.0691 & 8.4422 & 107.3256 & 17.3487 \\
        GPT-4o & 7.8806 & 1.1320 & 9.0126 & 113.9535 & 18.9988 \\
        Gemini 2.0 Flash & 7.9167 & 1.1374 & 9.0542 & 114.4186 & 19.0329 \\
        Claude 3.5 Haiku & 8.1220 & 1.1638 & 9.2858 & 117.5581 & 19.7633 \\
        Gemini 1.5 Pro & 8.7494 & 1.2470 & 9.9964 & 125.9302 & 21.6781 \\
        Llama 3.3 (70B) & 8.8747 & 1.2641 & 10.1387 & 128.6047 & 22.3635 \\
        GPT-4 Turbo & 8.9222 & 1.2700 & 10.1922 & 128.8372 & 22.3864 \\
        Gemini 2.0 Flash-Lite & 8.9279 & 1.2709 & 10.1988 & 129.0698 & 22.5757 \\
        Grok 2 & 10.1612 & 1.4328 & 11.5940 & 146.0465 & 26.1733 \\
        Llama 3.1 (70B) & 13.8681 & 1.9417 & 15.8099 & 197.7907 & 21.7054 \\
        \bottomrule
    \end{tabular}%
    }
\end{table}

\begin{table}[H]
    \caption{Performance and resource usage on \textbf{Two Pointers} problems from \textbf{Benchmark Set - II}.}
    \label{tab:llm-benchmark-set-ii-prb-twopointers}
    \centering
    \scriptsize
    \resizebox{\textwidth}{!}{%
    \begin{tabular}{lccccc}
        \toprule
        \textbf{Model} & \textbf{Avg. Pkg Energy (J)} & \textbf{Avg. RAM Energy (J)} & \textbf{Avg. Total Energy (J)} & \textbf{Avg. Runtime (ms)} & \textbf{Avg. Mem (MB·s)} \\
        \midrule
        \textbf{Canonical Solution} & \cellcolor{gray!25}5.9835 & \cellcolor{gray!25}0.8606 & \cellcolor{gray!25}6.8442 & \cellcolor{gray!25}86.6129 & \cellcolor{gray!25}16.5701 \\
        Grok 2 & 5.9560 & 0.8579 & 6.8139 & 85.6452 & 16.1937 \\
        DeepSeek v3 (37B) & 5.9782 & 0.8585 & 6.8368 & 85.4839 & 16.3254 \\
        Claude 3.5 Haiku & 5.9985 & 0.8610 & 6.8595 & 86.6129 & 16.5482 \\
        Gemini 1.5 Pro & 6.0013 & 0.8621 & 6.8634 & 86.2903 & 16.2923 \\
        Gemini 2.0 Flash & 6.0294 & 0.8673 & 6.8966 & 86.7742 & 16.4732 \\
        GPT-4o & 6.0713 & 0.8711 & 6.9424 & 87.0968 & 16.5650 \\
        Claude 3.5 Sonnet & 6.1439 & 0.8827 & 7.0266 & 88.3871 & 17.1383 \\
        Gemini 2.0 Flash-Lite & 6.1844 & 0.8869 & 7.0713 & 88.7097 & 17.3136 \\
        GPT-4 Turbo & 6.2097 & 0.8897 & 7.0994 & 89.0323 & 16.9718 \\
        Llama 3.1 (70B) & 6.3985 & 0.9147 & 7.3132 & 91.4516 & 18.4823 \\
        Llama 3.3 (70B) & 6.3977 & 0.9161 & 7.3139 & 91.7742 & 18.4822 \\
        \bottomrule
    \end{tabular}%
    }
\end{table}

\begin{table}[H]
    \caption{Performance and resource usage on \textbf{Sliding Window} problems from \textbf{Benchmark Set - II}.}
    \label{tab:llm-benchmark-set-ii-prb-slidingwindow}
    \centering
    \scriptsize
    \resizebox{\textwidth}{!}{%
    \begin{tabular}{lccccc}
        \toprule
        \textbf{Model} & \textbf{Avg. Pkg Energy (J)} & \textbf{Avg. RAM Energy (J)} & \textbf{Avg. Total Energy (J)} & \textbf{Avg. Runtime (ms)} & \textbf{Avg. Mem (MB·s)} \\
        \midrule
        \textbf{Canonical Solution} & \cellcolor{gray!25}5.2135 & \cellcolor{gray!25}0.7477 & \cellcolor{gray!25}5.9613 & \cellcolor{gray!25}74.5833 & \cellcolor{gray!25}13.2610 \\
        Claude 3.5 Haiku & 5.2219 & 0.7479 & 5.9698 & 75.4167 & 13.2504 \\
        Claude 3.5 Sonnet & 5.3706 & 0.7698 & 6.1404 & 76.8750 & 13.1980 \\
        DeepSeek v3 (37B) & 5.3906 & 0.7698 & 6.1604 & 77.2917 & 13.3448 \\
        GPT-4o & 5.4254 & 0.7744 & 6.1998 & 77.7083 & 13.3001 \\
        Gemini 1.5 Pro & 5.4296 & 0.7746 & 6.2042 & 77.9167 & 13.3303 \\
        Gemini 2.0 Flash-Lite & 5.4321 & 0.7763 & 6.2083 & 78.3333 & 13.2615 \\
        GPT-4 Turbo & 5.4715 & 0.7798 & 6.2513 & 77.7083 & 13.7686 \\
        Grok 2 & 5.5013 & 0.7850 & 6.2863 & 78.3333 & 13.2852 \\
        Llama 3.3 (70B) & 5.6110 & 0.8029 & 6.4140 & 80.8333 & 14.4033 \\
        Llama 3.1 (70B) & 5.6512 & 0.8065 & 6.4577 & 80.8333 & 14.3844 \\
        Gemini 2.0 Flash & 6.1148 & 0.8648 & 6.9796 & 87.0833 & 13.6598 \\
        \bottomrule
    \end{tabular}%
    }
\end{table}

\begin{table}[H]
    \caption{Performance and resource usage on \textbf{Bit Manipulation} problems from \textbf{Benchmark Set - II}.}
    \label{tab:llm-benchmark-set-ii-prb-bitmanipulation}
    \centering
    \scriptsize
    \resizebox{\textwidth}{!}{%
    \begin{tabular}{lccccc}
        \toprule
        \textbf{Model} & \textbf{Avg. Pkg Energy (J)} & \textbf{Avg. RAM Energy (J)} & \textbf{Avg. Total Energy (J)} & \textbf{Avg. Runtime (ms)} & \textbf{Avg. Mem (MB·s)} \\
        \midrule
        \textbf{Canonical Solution} & \cellcolor{gray!25}4.7403 & \cellcolor{gray!25}0.6727 & \cellcolor{gray!25}5.4130 & \cellcolor{gray!25}68.4848 & \cellcolor{gray!25}8.9917 \\
        GPT-4o & 5.1327 & 0.7223 & 5.8550 & 73.6364 & 9.0154 \\
        GPT-4 Turbo & 5.1361 & 0.7235 & 5.8595 & 73.7879 & 9.1655 \\
        Claude 3.5 Sonnet & 5.7236 & 0.8000 & 6.5236 & 81.6667 & 9.4019 \\
        Claude 3.5 Haiku & 6.1771 & 0.8589 & 7.0361 & 88.1818 & 9.3643 \\
        Gemini 2.0 Flash-Lite & 6.2344 & 0.8670 & 7.1014 & 89.3939 & 9.5377 \\
        DeepSeek v3 (37B) & 7.7945 & 1.0650 & 8.8595 & 109.8485 & 9.9943 \\
        Llama 3.1 (70B) & 9.7048 & 1.3150 & 11.0198 & 135.6061 & 10.7411 \\
        Grok 2 & 28.7344 & 3.7917 & 32.5261 & 394.6970 & 15.7563 \\
        Llama 3.3 (70B) & 29.1011 & 3.8211 & 32.9221 & 395.7576 & 30.5932 \\
        Gemini 2.0 Flash & 31.2664 & 4.0761 & 35.3424 & 424.5455 & 12.9024 \\
        Gemini 1.5 Pro & 38.0906 & 4.9780 & 43.0686 & 516.2121 & 15.7420 \\
        \bottomrule
    \end{tabular}%
    }
\end{table}

\begin{table}[H]
    \caption{Performance and resource usage on \textbf{Sorting} problems from \textbf{Benchmark Set - II}.}
    \label{tab:llm-benchmark-set-ii-prb-sorting}
    \centering
    \scriptsize
    \resizebox{\textwidth}{!}{%
    \begin{tabular}{lccccc}
        \toprule
        \textbf{Model} & \textbf{Avg. Pkg Energy (J)} & \textbf{Avg. RAM Energy (J)} & \textbf{Avg. Total Energy (J)} & \textbf{Avg. Runtime (ms)} & \textbf{Avg. Mem (MB·s)} \\
        \midrule
        \textbf{Canonical Solution} & \cellcolor{gray!25}5.2085 & \cellcolor{gray!25}0.7427 & \cellcolor{gray!25}5.9512 & \cellcolor{gray!25}75.2903 & \cellcolor{gray!25}9.9341 \\
        DeepSeek v3 (37B) & 5.2352 & 0.7472 & 5.9824 & 75.9355 & 9.8572 \\
        Gemini 2.0 Flash & 5.2943 & 0.7539 & 6.0483 & 76.5161 & 9.9365 \\
        Grok 2 & 5.3334 & 0.7601 & 6.0934 & 77.2258 & 9.8291 \\
        GPT-4o & 5.3683 & 0.7632 & 6.1315 & 77.8065 & 10.1040 \\
        Claude 3.5 Haiku & 5.3702 & 0.7627 & 6.1329 & 77.2258 & 10.1022 \\
        Gemini 1.5 Pro & 5.3963 & 0.7659 & 6.1622 & 77.7419 & 9.8676 \\
        Gemini 2.0 Flash-Lite & 5.5224 & 0.7842 & 6.3066 & 79.8065 & 10.3771 \\
        Claude 3.5 Sonnet & 5.8338 & 0.8249 & 6.6587 & 83.6774 & 10.4589 \\
        Llama 3.3 (70B) & 6.0050 & 0.8481 & 6.8531 & 86.4516 & 11.0881 \\
        Llama 3.1 (70B) & 6.9970 & 0.9767 & 7.9737 & 99.4839 & 11.3818 \\
        GPT-4 Turbo & 13.7246 & 1.8608 & 15.5854 & 192.6452 & 14.9146 \\
        \bottomrule
    \end{tabular}%
    }
\end{table}

\section{Code Analysis for Different LLMs}
\label{Appendix:F}

This section compares the code generated by different LLMs and analyzes the reasons behind their varying levels of energy efficiency compared to a canonical solution. The most energy-inefficient code is observed in algorithm categories such as Dynamic Programming, Backtracking, and Bit Manipulation. In these cases, the main cause of increased energy consumption is the absence of optimization techniques such as pruning. Without pruning, redundant computations are repeatedly executed, leading to higher computational load and memory usage. This, in turn, results in longer execution times and consequently higher energy consumption.

\subsection{Case I: Large Differences in Energy Consumption}

To evaluate this case, we take the solution generated for \textbf{LeetCode Problem 2305.} The problem details are as follows: \\
\textbf{Problem Name:} Fair Distribution of Cookies \\
\textbf{Algorithm Categories:} Dynamic Programming, Backtracking, Bit Manipulation \\
\textbf{Problem Description:} The task is to distribute the cookie bags as part of \texttt{cookies} list among \texttt{k} children as fairly as possible. The cookie bags should be distributed in such a way that all the cookies in the selected bag go to the same child. After distributing the cookies fairly, the program should return the minimum unfairness value which is the maximum number of cookies that a single child received during distribution. \\
\textbf{Input:} \texttt{cookies} \textit{(List of integers of length $n$, where $2 \leq n \leq 8$ and $1 \leq \texttt{cookies}[i] \leq 10^5$)} , \texttt{k} \textit{(An integer representing the number of children, where $2 \leq \texttt{k} \leq n$)} \\
\textbf{Output:} The minimum unfairness value \\

\begin{figure}[H]
    \centering
    \includegraphics[width=1\linewidth]{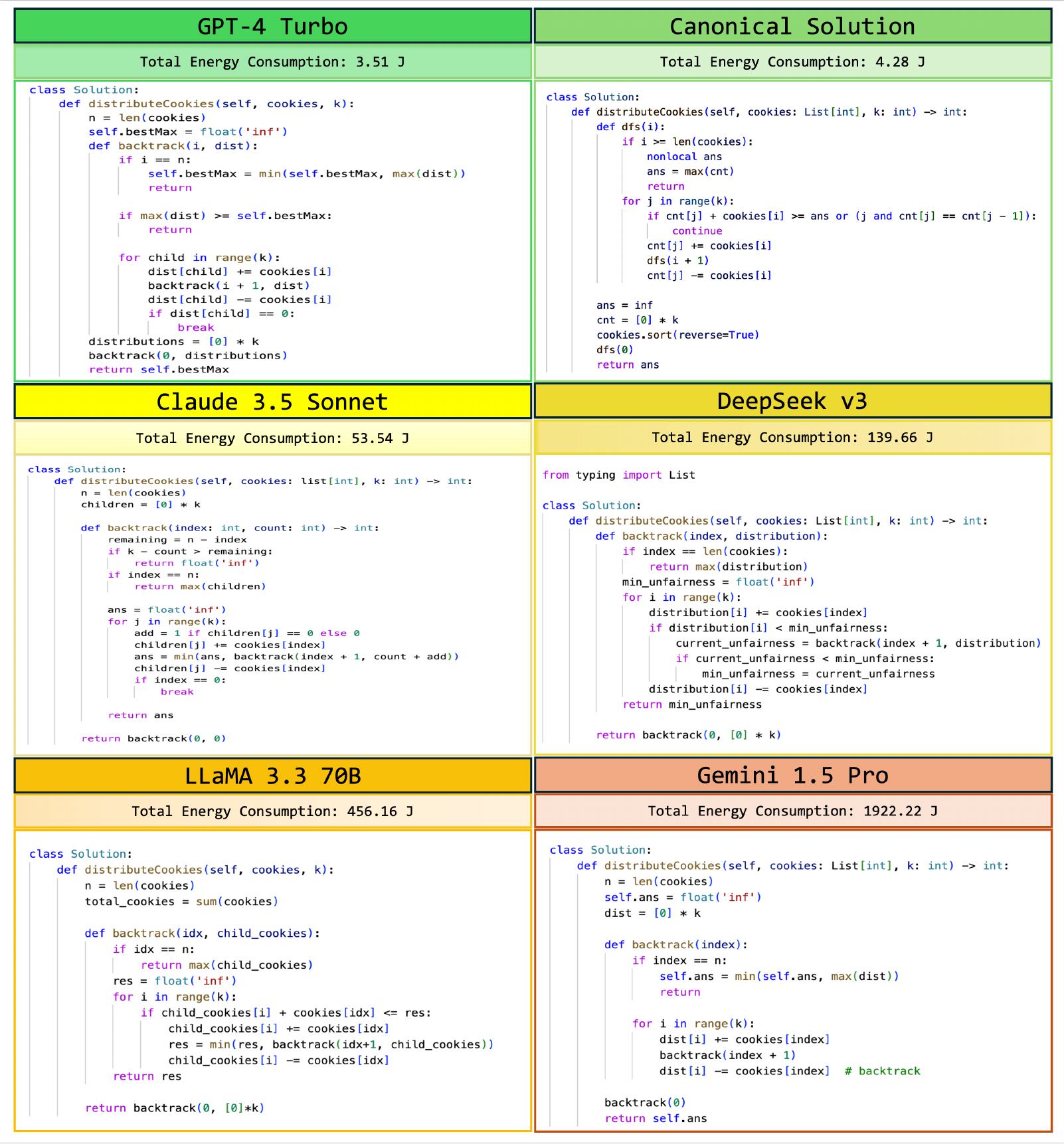}
    \caption{The above figure shows the canonical solution and the code generated by different LLMs along with their respective energy consumptions for LeetCode Problem 2305.}
    \label{fig:Code Analysis for Problem 2305}
\end{figure}

The following are our observations.
\begin{enumerate}
\item The \textbf{best solution} is the solution generated by LLMs \textbf{Claude Haiku} and \textbf{GPT-4 Turbo} as they apply effective pruning strategies to eliminate redundant calculations. 
\item The \textbf{Canonical Solution} has a \textbf{slightly higher} energy consumption than the best solution as it misses one pruning strategy of eliminating redundant calculation of symmetric paths.  
\item \textbf{GPT-4o} and \textbf{Gemini 2.0 Flash Lite} generate solutions that closely resemble the canonical version, and therefore demonstrate similar energy consumption profiles.
\item \textbf{Claude 3.5 Sonnet} produces a solution that has \textbf{15 times} higher energy consumption as compared to the best solution as it uses a minimum unfairness value that resets with each recursive call and lacks other checks that can reduce redundant computations. However, it does apply the pruning strategy to remove redundant calculation of symmetric paths.
\item \textbf{DeepSeek v3} produces a \textbf{slightly worse solution} than the canonical solution. This solution uses a minimum unfairness value that resets with each recursive call and lacks the symmetric path check, leading to more unnecessary computations, and thus resulting in energy consumption nearly \textbf{39 times} more than the best solution.
\item \textbf{LLaMA 3.3 70B} produces an \textbf{even worse solution} than \textbf{DeepSeek v3}. Although the logic produced by both LLMs is very similar, LLaMA 3.3 70B does more redundant calculations by evaluating the branches with the same values. This results in energy consumption which is \textbf{3 times} more than the DeepSeek v3 solution and \textbf{129 times} more than the best solution.
\item \textbf{Gemini 1.5 Pro} and \textbf{Gemini 2.0 Flash} generate the \textbf{worst solutions}. Both LLMs generate the same solution, which does not have any pruning logic in place, resulting in up to \textbf{450 times} the energy consumption of the\textbf{ canonical solution} and \textbf{548 times} the energy consumption of the \textbf{best solution}.
\end{enumerate}
Figure~\ref{fig:Code Analysis for Problem 2305} presents both the canonical solution and the representative outputs of selected LLMs, effectively highlighting the differences in energy consumption and supporting our performance analysis.

\subsection{Case II: Comparable Difference in Energy Consumption}
To evaluate this case, we take the solution generated for \textbf{LeetCode Problem 740.} The problem details are as follows: \\
\textbf{Problem Name:}  Delete and Earn \\
\textbf{Categories:} Dynamic Programming. \\ 
\textbf{Problem Description:} Given a list of integers \( \text{nums} \), the goal is to earn the maximum number of points by repeatedly performing the following operation: select any element \( \text{nums}[i] \) from the list, earn points equal to its value, and then remove that element along with all elements equal to \( \text{nums}[i] - 1 \) and \( \text{nums}[i] + 1 \) from the list. \\
\textbf{Input:} \texttt{nums} \textit{(List of integers containing the number of points where $1 \leq \text{nums.length} \leq 2 \times 10^{4}, 1 \leq \text{nums}[i] \leq 10^{4}$)}. \\
\textbf{Output:} The maximum number of points that can be earned. 

\begin{figure}[H]
    \centering
    \includegraphics[width=1\linewidth]{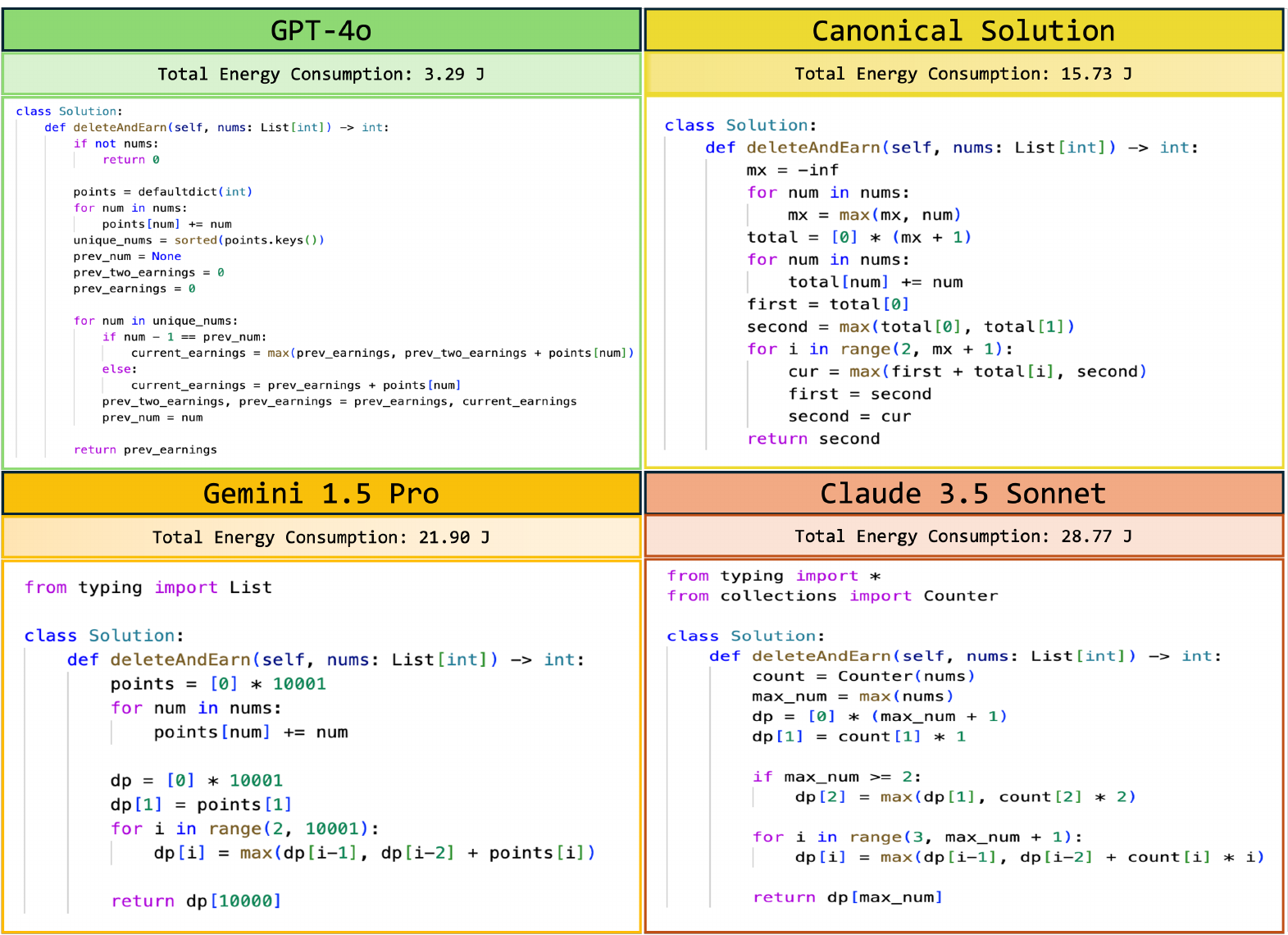}
    \caption{The above figure shows the canonical solution and the code generated by different LLMs along with their respective energy consumptions for LeetCode Problem 740.}
    \label{fig:Code Analysis for Problem 740}
\end{figure}

Figure~\ref{fig:Code Analysis for Problem 740} showcases the canonical solution to LeetCode Problem 740, along with representative outputs from selected LLMs, highlighting our comparative analysis.

The following are our observations.
\begin{enumerate}
\item The \textbf{best solution} is generated by \textbf{GPT-4o}, \textbf{Grok2}, \textbf{LLaMA 3.1 70B}, and \textbf{Gemini 2.0 Flash-Lite}. These models use an optimal approach using a few variables to iteratively compute the result.  Their solutions achieve a time complexity of \( \mathcal{O}(len(\text{nums})) \), resulting in the \textbf{lowest energy consumption} across all models evaluated.
\item The \textbf{Canonical solution} produces a solution that consumes \textbf{5 times} more energy than the best solution. This implementation relies on constructing a list with a size proportional to the maximum value in the input array, resulting in a time complexity of \( \mathcal{O}(\max(\text{nums})) \). This increased both memory usage and computational cost. 
\item \textbf{Gemini 1.5 Pro} produces a solution with energy consumption roughly \textbf{7 times} greater than the best solution. Similar to the canonical approach, it uses a list-based method to accumulate points. However, the list is statically sized to a fixed upper bound of 10,000, irrespective of the actual maximum value in \text{nums}. This leads to a suboptimal time complexity of \( \mathcal{O}(10^4) \).
\item The \textbf{worst solution} is produced by \textbf{Claude-Sonnet} and \textbf{LLaMA 3.1 70B}. While structurally similar to the canonical approach in utilizing a list indexed by element values, their implementations introduce additional redundant multiplications during the computation. These unnecessary operations further increase computational overhead, resulting in energy consumption nearly \textbf{9 times} higher than the best solution.
\end{enumerate}




\end{document}